\documentclass[]{aastex631}
\usepackage{amsmath}

\newcommand{\U}[1]{\underline{#1}} 
\shorttitle{Jupiter's Interior}
\shortauthors{Militzer and Hubbard}
\graphicspath{{./}{figures/}}

\begin{document}

\title{Study of Jupiter's Interior: Comparison of 2, 3, 4, 5, and 6 Layer Models}

\author{Burkhard Militzer}
\affil{Department of Earth and Planetary Science, Department of Astronomy,\\ University of
California, Berkeley, CA, 94720, USA}

\author{William B. Hubbard}
\affil{Lunar and Planetary Laboratory, The University of
Arizona, Tucson, AZ 85721, USA}



\begin{abstract}
With the goal of matching spacecraft measurements from {\em Juno} and {\em Galileo} missions, we construct ensembles of 2, 3, 4, 5, and 6 layer models for Jupiter's interior. All except our two layer models can match the planet's gravity field as measured by the {\em Juno} spacecraft. We find, however, that some model types are more plausible than others. In the best three layer models, for example, the transition from molecular to metallic hydrogen needs to be at $\sim$500 GPa while theory and experiments place this transition at $\sim$100 GPa. Four layer models with a single sharp boundary between core and mantle would be short-lived due to rapid convective core erosion. For this reason, we favor our five layer models that include a dilute core surrounded by a stably stratified core transition layer. Six layer models with a small compact core are also possible but with an upper limit of 3 Earth masses for such a compact core. 
All models assume a 1 bar temperature of 166.1 K, employ physical equations of state, and are constructed with the nonperturbative Concentric Maclaurin Spheroid (CMS) method. We analyze the convergence of this method and describe technical steps that are needed to make this technique so efficient that ensembles of models can be generated. 
\end{abstract}



\section{Introduction} \label{sec:intro}

The space missions {\em Juno} \citep{Bolton2017} and {\em Cassini} \citep{Spilker2019} have provided us with a wealth of new data for Jupiter and Saturn, the two largest planets in our solar system. Multiple close flybys have determined the gravity fields of these planets with exquisite precision, far exceeding the earlier measurements by the {\em Pioneer} and {\em Voyager} missions~\citep{CS85,CA89}. The earlier missions determined Jupiter gravity harmonics to be $J_2=14697 \pm 1$, $J_4=-584 \pm 5 $, and $J_6=31 \pm 20 $ (all multiplied by 10$^6$ and adjusted to an equatorial radius of 71492 km) while today these coefficients are known to three orders of magnitude higher precision: $J_2=14696.5735 \pm 0.0017$, $J_4=‐586.6085 \pm 0.0024 $, and $J_6=34.2007 \pm 0.0067$~\citep{Durante2020}. This improvement has led to revision in the models of Jupiter's interior required to match these gravity measurements. When interior models were only constrained by Jupiter's mass, equatorial radius, $J_2$ and $J_4$, traditional three layer models for Jupiter's interior sufficed~\citep{Stevenson1982,Guillot2005}. These models typically included a compact core comprising up to 100\% of heavy elements, presumably left over from the planet's formation by core accretion~\citep{Bodenheimer1986}. All these three layer models assumed Jupiter's envelope to be inhomogeneous~\citep{saumon-apj-04,militzer-apj-08,Nettelmann2012}, introducing a discontinuity in the helium or in the heavy elements fraction at some radius or pressure in the envelope. The motivation for the discontinuity was in part empirical because one needed an additional free parameter to match $J_4$ while adjusting the core mass and heavy element fraction in the envelope enabled one to match the planet's mass and $J_2$. An exception were the homogeneous models reported by \citet{militzer-apj-08} who invoked deep differential rotation to match $J_4$ instead. However, the required wind depth of $\sim$10$\,$000 km far exceeded values implied by Jupiter's odd gravity harmonics~\citep{Kaspi2018} and typical magnetic field models~\citep{Christensen2020}. Both these latter approaches favor a wind depth of $\sim$3000 km although an agreement has yet to be reached on precisely how the wind speeds decay with depth.

A variety of model assumptions and techniques have been invoked to model Jupiter's interior structure in the past. \citet{Ni2018} combined the theory of figures \citep{ZT1978} with the approach by \citet{AndersonSchubert2007} to adjust coefficients of an empirical polynomial function that represents the Jupiter's interior density profile in order to match the {\em Juno} gravity measurements. \citet{Nettelmann2021} extended the ToF to technique to seventh order. \citet{MilitzerHubbard_MOI_2023} constructed physical and abstract models of Jupiter's interior structure to demonstrate that the gravity coefficients $J_2$, $J_4$ and $J_6$ are sufficient to constrain Jupiter's moment of inertia very tightly. The physical models rely on realistic equations of state and assumptions for the compositions while abstract models represent the planet's interior by a small number of constant-density. \citet{Howard2023} studied the effects of various published equation of state (EOS) tables and also introduced an EOS modification. \citet{Moll2017}, \citet{Muller2020}, and \citet{Helled_2022} studied the Jupiter's interior convection and the evolution of a primordial, compact core that was originally composed to 100\% of heavy elements. \citet{Liu2019} investigated whether such a core could be diluted by a giant impact. Core dilution is plausible because {\em ab initio} computer simulations have shown that all typical core materials such as water, silicates and iron are soluble in metallic hydrogen at megabar pressures~\citep{WilsonMilitzer2012,WilsonMilitzer2012b,Wahl2013,Gonzalez2014}.

\citet{Kerley2004} first constructed a homogeneous baseline model to demonstrate that the pre-{\em Juno} gravity data cannot be matched unless the envelope is represented by at least two distinct layers. The need for Jupiter's envelope to be inhomogeneous was recently reconfirmed by \citet{Miguel2022}. The original justification for the inhomogeneity was a first-order phase transition from molecular to metallic hydrogen. However, {\em ab initio} computer simulations~\citep{Vorberger2010,Vorberger2012} have since shown that such a transition likely exists only at low temperature in hydrogen~\citep{Morales2010} while the temperatures in giant planet interiors are everywhere above the critical point so that molecular dissociation there occurs gradually. So this process by itself is insufficient to cause an inhomogeneity. Alternatively, the phase separation of hydrogen-helium mixtures~\citep{Morales2013} has been considered but so far there is insufficient evidence that the falling helium droplets will sequester any heavy elements besides neon~\citep{WilsonMilitzer2010}. 

In this article, we incorporate the nonperturbative concentric Maclaurin spheroid (CMS) method~\citep{Hubbard2013} into a streamlined modeling procedure which achieves the high precision needed to exploit {\it Juno} constraints, while automatically matching basic constraints such as mass, equatorial radius, and low-degree zonal gravity harmonics. With Monte Carlo sampling~\citep{Militzer_QMC_2023} we then efficiently generate large numbers of models with increasing degrees of complexity. We construct models for Jupiter's interior assuming different compositions and layered structures. We invoke one consistent set of physical assumptions. We adopt one equation of state derived with {\em ab initio} simulations~\citep{MH13} without introducing any corrections to lower its density, nor do we vary the temperature at 1 bar, 166.1~K measured by the {\em Galileo} entry probe ~\citep{Niemann1998,Wong2004}. This temperature value is also consistent with earlier radio occultation measurements by the {\em Voyager} spacecraft that yielded a value of 165~$\pm$~5~K~\citep{Lindal1981}. These remote observations very recently re-analyzed by \citet{Gupta2022} who determined higher temperatures of 167$\pm$4 and 170$\pm$4~K for latitudes of 6$^\circ$S and 12$^\circ$N respectively. 

\begin{figure}
\gridline{
          \fig{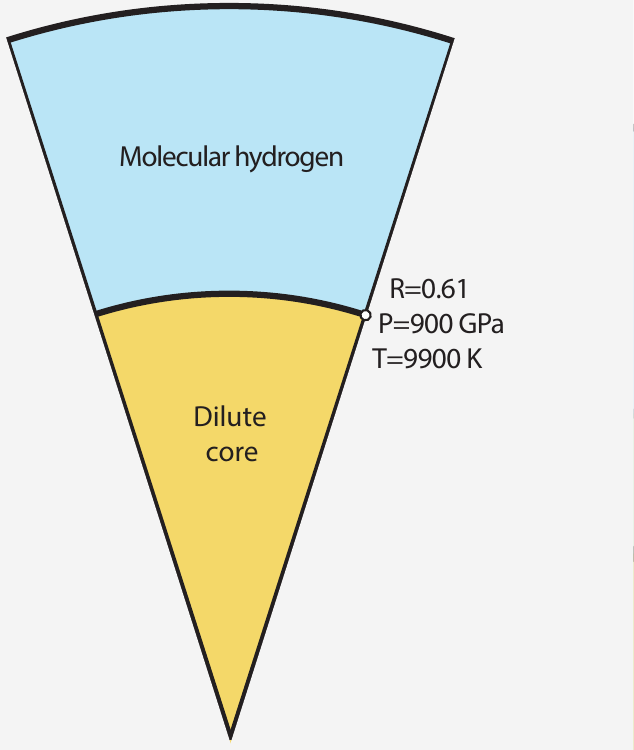}{0.20\textheight}{Two layer model}
          \fig{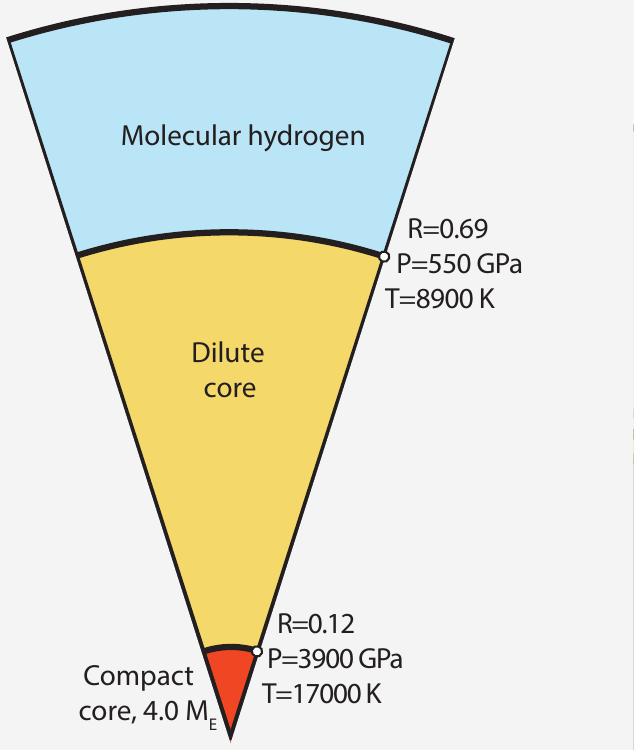}{0.20\textheight}{Three layer model}
          \fig{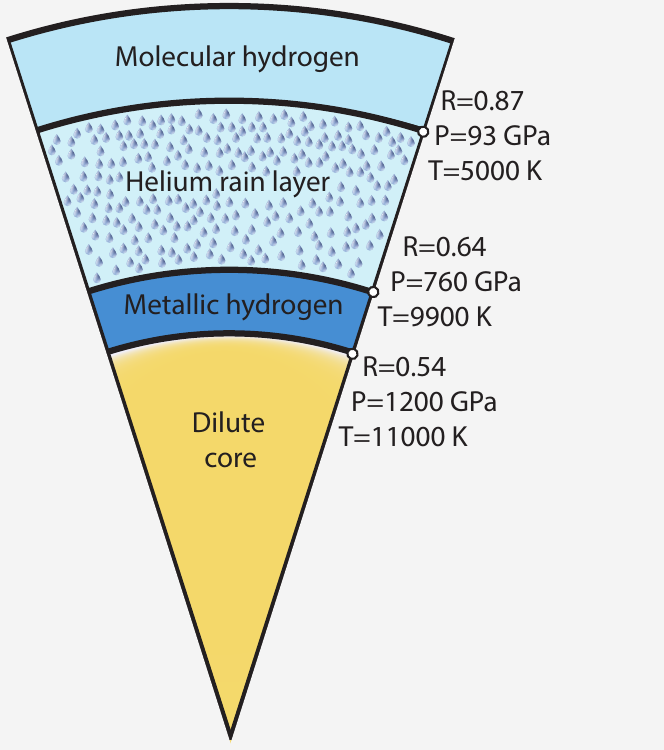}{0.21\textheight}{Four layer model (A)}
          }
\gridline{
          \fig{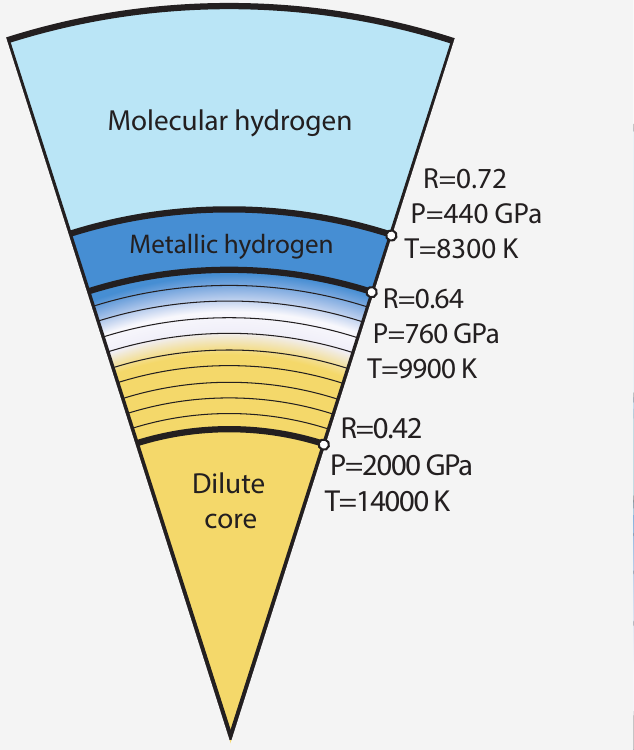}{0.19\textheight}{Four layer model (B)}
          \fig{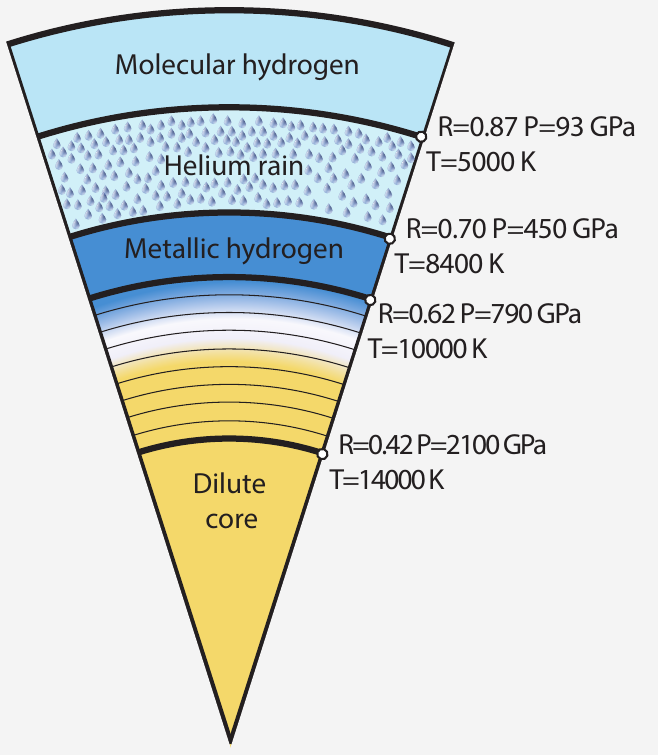}{0.20\textheight}{Reference model with five layers (A)}
          \fig{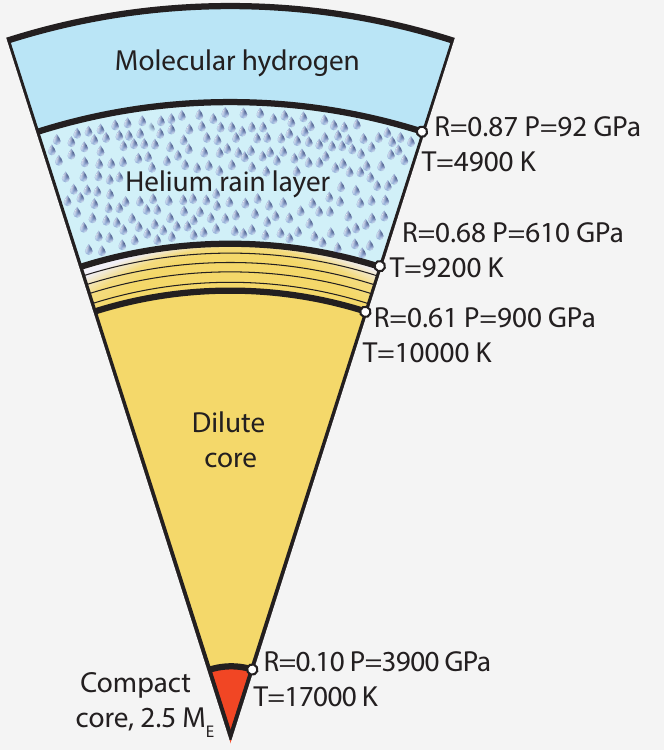}{0.20\textheight}{Five layer model (B)}
          }
\gridline{
          \fig{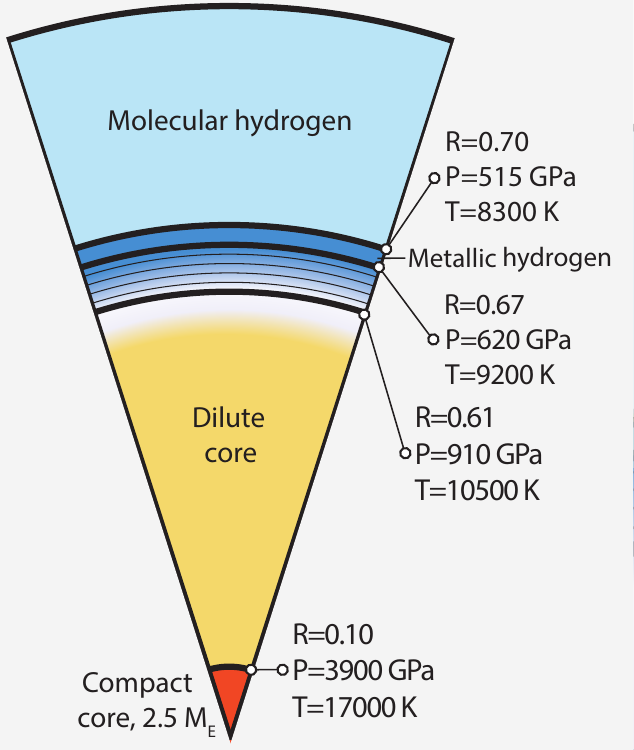}{0.19\textheight}{Five layer model (C)}
          \fig{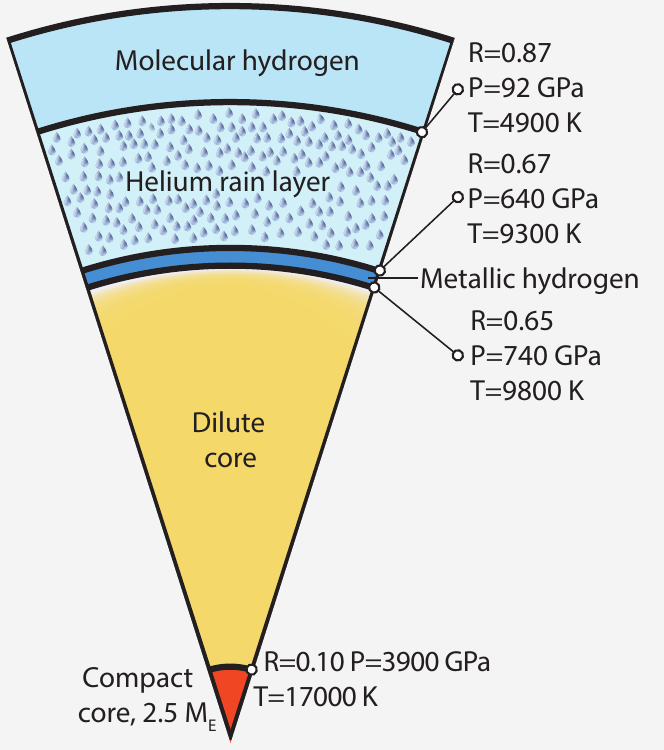}{0.20\textheight}{Five layer model (D)}
          \fig{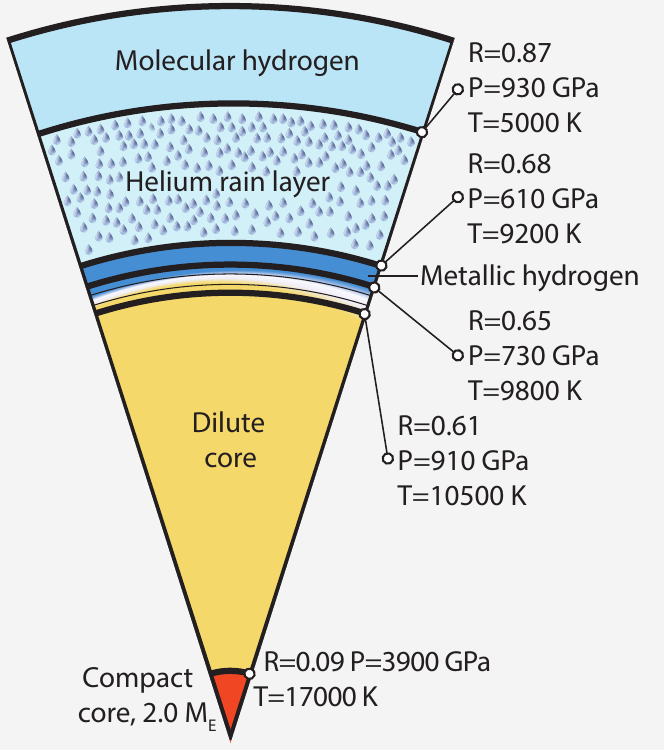}{0.20\textheight}{Six layer model}
          }          
\caption{Illustration of our 2, 3, 4, 5 and 6 layer models in Tab.~\ref{tab:23456}. At every layer boundary we specify the pressure, temperature, and fractional volumetric radius. The collection of thin solid lines represent the core transition layer that may surround the dilute core. In the center is our reference model with five layers. The two types of four layer models are generated by removing either the helium rain or the core transition layers. The five layer models B, C, and D were obtained by removing one layer one from the reference model while adding a compact core. Such a core was included in our final six layer model. It also distinguishes our simpler two and three layer models.
\label{fig:23456}}
\end{figure}

We construct models that have between two and six layers as we illustrate in Fig.~\ref{fig:23456}. We start from a reference model with five layers \citep{DiluteCore} that contain an outer layer of molecular hydrogen, a helium rain layer, a layer of metallic hydrogen, a core transition layer, and a prominent dilute core in the center that may extend up to 60\% of the planet's radius. We generate two types of four layer models by either eliminating the helium rain layer or the core transition layer from our five layer reference models. Six layer models are derived by inserting a compact core into the reference models. Under our model assumptions, we will demonstrate that such a compact core cannot be heavier than 3 Earth masses for the models to be compatible with Jupiter's gravity measurements. We also generate alternate types of five layer models by removing either the helium rain layer,  metallic hydrogen layer, or the core transition layer. 

Finally we revisit the traditional three layer models and even try to construct two layer models with just a core and an homogeneous envelope. These two layer models do not fit the gravity field, while our three, four, five, and six layer models all do. Still some of these models are more plausible than others. For example, matching the gravity field with three layer models requires us to place a transition between a helium-depleted outer layer and an helium-enriched inner layer at a pressure of 500 GPa, which is incompatible with the physics of hydrogen-helium mixtures~\citep{Morales2013}.

%
%

This paper is organized as follows. In section~\ref{sec:method}, we describe our model assumptions and explain how we efficiently converge on model solutions for the interior structure and winds that match the planet's radius, mass, gravity harmonic $J_2$, and protosolar helium abundance. In section~\ref{sec:results}, we discuss the results for the different model types. We show that two layer models are unable to match $J_4$ and explain why three layer models need to have a molecular-to-metallic transition at $\sim $500~GPa. We then analyze the properties of 4, 5, and 6 layer models. We describe how we place constraints on the mass of the compact core and the helium distribution within the helium rain layer. We conclude with section~\ref{sec:conclusions}.

\section{Methods and Model Assumptions} \label{sec:method}

\subsection{Methods for the Interior}

In this paper, we employ the CMS method to construct models of rapidly rotating giant planets that meet certain observational constraints such as planet mass, equatorial radius, helium content, etc. In principle one has two choices of how to meet such constraints because the CMS method essentially provides a way to converge onto a hydrostatic solution for a given pressure-density relationship by adjusting the shapes of all spheroids. One can then wait until the CMS method has converged, evaluate how well the constraints have been met, then vary certain model parameters, and restart the CMS procedure. Alternatively, one can adjust certain model parameters during the CMS convergence procedure so that certain criteria are met automatically as a hydrostatic solution is found. The latter method is obviously much more efficient because it typically does not increase the cost of the CMS procedure and more importantly, it reduces the number of dimensions of the parameter space that model optimization and Monte Carlo calculation need to explore. We now discuss how we converge onto a valid planetary interior model that satisfies a number of plausible constraints:

(1) For that reason, almost all CMS calculations fix the equatorial radius $a$, assumed to have been measured with such high precision that its remaining uncertainty has a negligible effect on all computed planet properties. Here we use $a=71492$~km (n.b.~the error bar on $a$ is $\pm 4$~km) and associate it with a pressure of 1 bar. (Conversely, in \citet{Wahl2021}, the equatorial radius was varied during the CMS convergence in order to match the planet mass as models of specific exoplanets with different rotation periods and oblatenesses were constructed.)

(2a) Next we adjust the helium concentration in the interior so that the planet overall matches the protosolar hydrogen-helium fraction of $\tilde{Y}_0 \equiv Y_0/(X_0+Y_0)=0.2777$ derived from $X_0=0.7112$ and $Y_0=0.2735$~\citep{Lodders2010}. In the outer molecular layer, we set $Y_1/(X_1+Y_1)=0.2369$ to be consistent with measurements of the Galileo entry probe, $Y/(X+Y) = 0.238 \pm 0.005$~\citep{vonzahn-jgr-98}. We assume the helium mass fraction $\tilde{Y}_1 \equiv Y_1/(X_1+Y_1)$ remains constant in the molecular layer up to a pressure $P_{\rm rain,1}$ and from $P_{\rm rain,2}$ on, it is again constant at $\tilde{Y}_2 \equiv Y_2/(X_2+Y_2)$. In the helium rain layer from $P_{\rm rain,1}$ to $P_{\rm rain,2}$, we gradually switch from one value to the other as a function of $\log(P)$,

\begin{eqnarray}
\tilde{Y}(P) &=& \tilde{Y}_1 + F_2(x) \left[\tilde{Y}_2-\tilde{Y}_1 \right]\\
F_2(x) &=& x^\alpha\label{exp}\\
 x &=& \frac{\log{P} - \log{P_{\rm rain,1}}}{\log{P_{\rm rain,2}} - \log{P_{\rm rain,1}}}
\end{eqnarray}

The switching function, $F_2$, represents the fraction of $\tilde{Y}_2$ in the mixture and satisfies the expected values at the boundaries, $F_2(P=P_{\rm rain,1})=0$ and $F_2(P=P_{\rm rain,2})=1$. Besides representing a linear change from $\tilde{Y}_1$ to $\tilde{Y}_2$ when $\alpha=1$, it introduces the option of having a large helium fraction from that layer being sequestered by setting $\alpha \gg 1$ or increasing the helium fraction in that layer by specifying $\alpha \ll 1$. The simple form of Eq.~\ref{exp} has the advantage that the average fraction across the helium rain layer is given by,

\begin{equation}
\left< F_2 \right> = \int_0^1 dx \, F_2(x) = 1/(1+\alpha) \;\;.
\end{equation}

(2b) Next we construct a spline function to represent the cumulative mass as a function of $\log(P)$ so that we can derive the mass fraction that is stored in the layer between two pressure values. The global helium fraction, $\left< \right. \tilde{Y} \left. \right>$, is given by

\begin{eqnarray}
\left< \right. \tilde{Y} \left. \right> &=& m_1 \tilde{Y}_1 \;\;+\;\; m_{\rm rain} \tilde{Y}_{\rm rain} \;\;+\;\; m_2 \tilde{Y}_2\;,\\
&=& \left[m_1 + m_{\rm rain}(1-\left<F_2\right>) \right] \; \tilde{Y}_1 \;\;+\;\; \left[m_{\rm rain}\left<F_2\right> + m_2\right]\;\tilde{Y}_2\;,\label{eq:Y2}
\end{eqnarray}
where $m_1$, $m_{\rm rain}$, and $m_{2}$ represent the fractional masses of the three respective layers without the contributions from heavy elements. Eq.~\ref{eq:Y2} enables us to adjust $\tilde{Y}_2$ so that the global helium abundance matches the protosolar value, $\left< \right. \tilde{Y} \left. \right> = \tilde{Y}_0$.

(3) Finally, we discuss how the CMS procedure can converge directly onto the exact mass and $J_2$ value of the planet. We decided to match $J_2=(14696.5063 \pm 0.0006) \times 10^{-6}$ exactly because it has been measured with exquisite precision. (As recommend by \citet{Durante2020}, we do not include the tidal effects on $J_2$, estimated to be 6.72$\times$10$^{-8}$.)
The planet mass has previously been matched by introducing an overall density scaling factor $\beta$~\citep{MH13} or by adjusting the density of a compact core represented by a single innermost spheroid~\citep{Wahl2017a}. Since our current set of models do not have a compact core or its mass is very small, it is no longer practical to match the planet's mass by adjusting the core mass. So we instead match the planet's mass and $J_2$ by simultaneously adjusting $Z_1$ and $Z_2$ as the CMS method converges to a hydrostatic solution. $Z_1$ is the mass fraction of heavy elements in the outer layers of the planet that we assume to be constant from the atmosphere to the beginning of the core transition layer. Here our assumptions differ from those by \citet{Debras2019} who matched $J_4$ and $J_6$ by assuming Jupiter's outer enveloped is not fully convective and that the concentration of heavy elements decreased with increasing pressure. In this work, $Z_2$ is the mass fraction of heavy elements in the dilute core. In the core transition layer, we assume $Z$ to increase linearly as a function of $\log(P)$ from $Z_1$ and $Z_2$. 

We find that adjusting $Z_1$ and $Z_2$ to match $M$ and $J_2$ is much more efficient than first adjusting $Z_1$ to match the mass in an inner loop and then adjusting $Z_2$ to match $J_2$ in an outer loop. For example if a bisection procedure or Brent's method~\citep{numerical_recipes} are employed as an outer loop to bring the deviation $f(Z_2) = J_2(Z_2)-J_2^{\rm target}$ to zero, one has to converge multiple CMS models, which is inefficient. To start, one has to evaluate $f(Z_2)$ for two different values of $Z_2$ that yield $f$ values of opposite signs before $Z_2$ can be refined iteratively. This means that even if one already has a reasonably good starting value for $Z_2$, one has to step away from it and construct another hydrostatic CMS solution for a second $Z_2$ value that is ultimately not needed. This is unnecessary and inefficient. To avoid such unnecessary steps, we define two following functions,
\begin{eqnarray}
f_1(Z_1,Z_2) &\equiv& M(Z_1,Z_2) \;\;\;\;\; \;\;\;\;\; \;\;\;\;\; \;\;\;\;\; \,- M_{\rm planet}\\
f_2(Z_1,Z_2) &\equiv& M(Z_1,Z_2) \times J_2(Z_1,Z_2) - M_{\rm planet} \times J_2^{\rm target}
\end{eqnarray}
and employ the Newton-Raphson method to adjust $Z_1$ and $Z_2$ while the CMS method converges to a hydrostatic solution that then matches $M$ and $J_2$ automatically. This approach requires us to compute the derivatives of $M$ and $(MJ_2)$ with respect to $Z_1$ and $Z_2$, which we will explain in the following equations. (Alternatively, one may define $f_2(Z_1,Z_2) \equiv J_2(Z_1,Z_2) - J_2^{\rm target}$, which leads to iteration equations of similar complexity.)

In rare cases, the Jacobian derivative matrix (see. Eq.~\ref{Jac}) for a particular Newton-Raphson step cannot be inverted or the Newton-Raphson method starts oscillating between two states. If this happens we insert a 2D {\it regula falsi} step to depart from such a pathological situation. Details of our {\it regula falsi} method are given in appendix~\ref{RootFinder}. We find that this combined approach works very well except in cases when $P_{\rm rain,1}$ and $P_{\rm rain,2}$ exceed the central pressure of the planet and the variable $Z_2$ becomes irrelevant so that we no longer have two independent variables to match $M$ and $J_2$ simultaneously.

Within the CMS method~\citep{Hubbard2013}, the planet's interior is represented by a series of nested spheroids indexed by $j$, each with volume $V_j$. The total mass is derived from,
\begin{eqnarray}
M &=& \sum_j \rho_j (V_j-V_{j+1}) = \sum_j \delta_j V_j \;, \label{eq:M}\\
V_j &=& \frac{2\pi}{3} \lambda_j^3 \int_{-1}^{+1} d\mu  \; \zeta_j^3(\mu) \;, 
\end{eqnarray}
where $\delta_j=\rho_j - \rho_{j-1}$ represents the density difference of spheroid $j$ and next outer spheroid $j-1$. Following~\citet{Hubbard2013}, the shape function, $\zeta_j(\mu)\equiv \zeta_{j,k}$, is represented by a series of discrete quadrature points, $\mu_k=\cos(\theta_k)$, where $\theta_k$ are colatitude values. The distance of the spheroid surface from the origin is given by $r(\mu) = \lambda_j \zeta_j(\mu)$ where $\lambda_j$ is the equatorial radius of spheroid $j$.

In a similar fashion, one can derive an expression for the product $(M \times J_n)$,
\begin{eqnarray}
(MJ_n) &=& \sum_j \delta_j \hat{J}_{j,n} = \sum_j \rho_j \left( \hat{J}_{j,n}-\hat{J}_{j+1,n} \right) \label{eq:MJ}\\
\hat{J}_{j,n} &=& \frac{-2 \pi}{3+n}\; \lambda_j^{n+3} \int_{-1}^{+1} d\mu  \; P_n(\mu) \; \zeta_j^{n+3}(\mu) \;.
\end{eqnarray}
The coefficients $\tilde{J}_{j,n}$ from \citet{Hubbard2013} are recovered through, $\tilde{J}_{j,n} = \delta_j \hat{J}_{j,n} / (\lambda_j^n M) $. The conventional gravity harmonics are obtained by
\begin{equation}
J_n = \frac{1}{M} \sum_j \delta_j \hat{J}_{j,n} = \sum_j \lambda_j^n \tilde{J}_{j,n} \;. \label{weights}
\end{equation}
Equations~\ref{eq:M} and \ref{eq:MJ} would enable us to determine the derivatives of $f_1$ and $f_2$ with respect to $Z_1$ and $Z_2$ if we knew how the density, $\rho(\tilde{Y},Z)$, varied with the contents of helium and heavy elements. Following~\cite{HubbardMilitzer2016}, we invoke the additive volume rule to incorporate the heavy elements into our equation of state~\citep{MH13,Militzer2013}, $\rho^*$, that was computed for a single helium fraction, $Y^*$, without any heavy elements ($Y^*\equiv \tilde{Y}^*$),
\begin{equation}
\frac{\rho^*}{\rho(\tilde{Y},Z)} = (1-Z)A + Z \frac{\rho^*}{\rho_Z} \;\; \;\; \rm{with} \;\;\;\; A= \frac{ (1-\tilde{Y}) + (\tilde{Y}-Y^*) \frac{\rho^*}{\rho_{\rm He}} }{1-Y^*}  \label{eq:rho}
\quad.
\end{equation}
$\rho_{\rm He}$ is an EOS for pure helium~\citep{Mi06,Mi09,FPEOS} that enables us to perturb the helium fraction from our standard value, $Y^*$. We incorporate heavy elements in the envelope by following \citet{HubbardMilitzer2016} who conducted {\it ab initio} computer simulations of a solar-proportion mixture of H$_2$O:CH$_4$:NH$_3$ and then represented the results by a function $\rho_Z/\rho^*$ with that varies slowly pressure, which was found to be in agreement with later work by \citet{Soubiran2016}.

Equation~\ref{eq:rho} not only allows us to compute, $\rho(\tilde{Y},Z)$, it also makes the evaluation of $\frac{d\rho(\tilde{Y},Z)}{dZ}$ straightforward. It allows us to determine how the density of every layer $\rho_j$ in Eqs.~\ref{eq:M} and \ref{eq:MJ} depends on $Z$ and therefore gives us access to the derivatives of $f_1$ and $f_2$ that we need when we employ the Newton-Raphson iterations to match the planet's mass and $J_2$ by adjusting $Z_1$ and $Z_2$ as the CMS procedure converges to a hydrostatic solution. 

\begin{figure}
\gridline{\fig{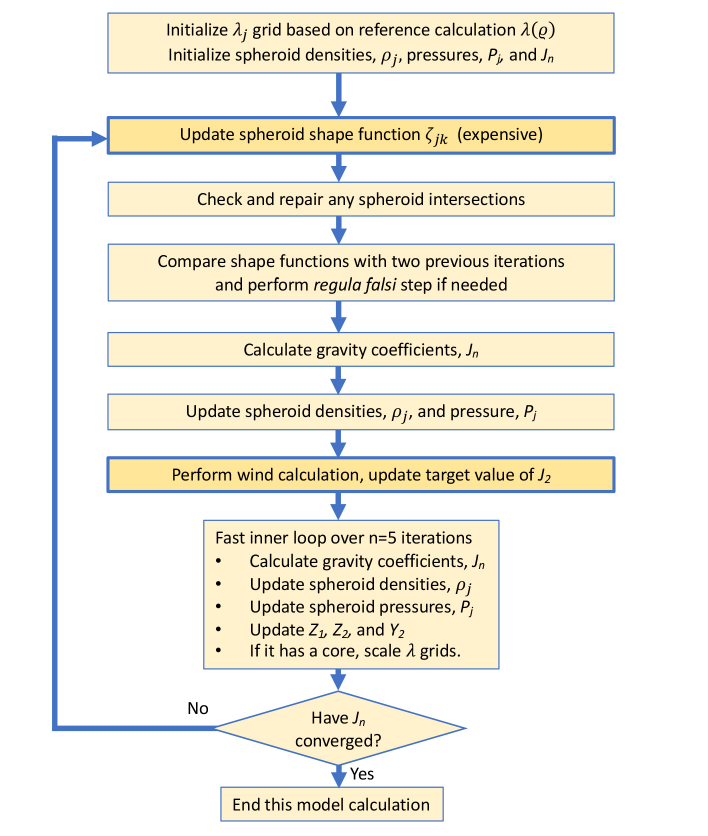}{0.7\textwidth}{}}
\caption{
Our preferred technique to efficiently converge onto a selfconsistent solution that includes a hydrostatic description of the interior with the CMS method and a compatible wind model. It is novel that the planet's mass, $J_2$, and global helium abundance are matched automatically as the calculation converges without a large computational overhead. Calculations of spheroid shapes and the wind contributions to the $J_n$ are performed only once in the main loop because they are rather expensive. In the inner loop, the gravity coefficients, densities, pressures, the helium fraction $Y_2$ and heavy element abundances $Z_1$ and $Z_2$ are updated more often because they scale linearly with spheroid number. If the model includes a compact core of a specific mass target, the $\lambda$ grid is updated in the inner loop to match that. Besides converging the gravity coefficients, $J_n$, other criteria for mass, $J_2$, or the overall helium abundance may be employed to terminate the main loop.
\label{fig:conv}}
\end{figure}

In Fig.~\ref{fig:conv}, we describe our preferred way to efficiently converge a CMS calculation. We start by setting up an optimized $\lambda$ grid~\citep{MilitzerSaturn2019} that anchors the equatorial radii of all spheroids. All other spheroid points are adjusted so that an equipotential emerges. Based on an earlier CMS calculation for Jupiter, the $\lambda$ grid is chosen so that the corresponding density values fall on a logarithmic grid such that $\rho(\lambda_j) / \rho(\lambda_{j+1}) = $constant. This was shown to be a good choice \citep{MilitzerSaturn2019} and CMS results do not depend on the specifics of the earlier Jupiter model and there is no need to update the $\lambda$ grid iteratively. 

At the beginning of the main loop in Fig.~\ref{fig:conv}, we use Newton's method to update the shapes of all spheroids $\zeta_{j,k}$ \citep{Hubbard2013}, which is the most time consuming step besides the wind calculations. If the shapes are far from equilibrium, it may happen that the updated shapes of two adjacent spheroids cross, which is unphysical. We detect such rare events and correct spheroid shapes. It may also happen that the shapes start oscillating between two states. When we detect this, we insert a {\it regula falsi} step.

We perform the thermal wind calculations once in the main loop to update the target value for $J_2 = J_2^{\rm Juno} - J_2^{\rm wind}$.
In the fast, inner loop we update the gravity coefficients, $J_n$, spheroid densities $\rho_j$ and pressures, $P_j$, because these calculations are inexpensive and scale linearly with spheroid number. We also update $Z_1$, $Z_2$, and $Y_2$ as described above. 

If the model includes a compact core of prescribed mass, some additional steps are needed. First we derive a scale factor, $f= (m_{\rm core}^{\rm current} / m_{\rm core}^{\rm target})^{-1/3}$ and then scale the $\lambda$ grid inside the core region with $\lambda_j = f \times r_{\rm core} \times (N_l - j)/N_l^{\rm core}$ so that the equatorial radius of the outermost core spheroid becomes $r_{\rm core}^{\rm (new)}=f \times r_{\rm core}^{\rm
  (old)}$. $N_l^{\rm core}$ is the number of spheroids in the core while $N_l$ is the total spheroid number. We also update the logarithmic $\lambda$ grid in the envelope so that it ends at the revised $r_{\rm core}^{\rm (new)}$ value.

We find that our model predictions are not very sensitive to the assumed composition of a compact core because the density of typical core materials is much higher than that of hydrogen and helium, so it is primarily the mass of a compact core that matters. We still see some deviations when we compare models with a rocky compact core with models that assume a 50:50 rock-ice mixture. We follow \citet{Seager07} when we adopt equations of state for the core materials but then complement it with more results from more recent {\it ab initio} simulations \citep{MilitzerWilson2010,WilsonMilitzer2014,GonzalezMilitzer2023}. For rock, we assume a terrestrial mixture of MgSiO$_3$ and iron with an iron mass fraction of 32.5\%. We represent the ice in the compact core by assuming a pure H$_2$O composition. 

\subsection{Assumptions for Model Ensembles}
\label{sec:model_assumptions}

We employ Monte Carlo methods~\citep{Militzer_QMC_2023,QMC_code} to construct ensembles of Jupiter models by accepting and rejecting moves according to the $\exp(-\chi^2/2)$ function that combines the terms, $\chi^2 = \chi^2_J + \chi^2_{\rm H-He} + \chi^2_{\rm wind} + \chi^2_{\rm guide}$. The first and most important term measures the deviations of even and odd gravity harmonics between model predictions and the {\em Juno} measurements~\citep{Durante2020},
\begin{equation}
  \chi^2_J = \sum\limits_{i=1}^{10} \left[ \frac{ J_{i}^{\rm model}  - J_{i}^{\rm Juno} }{ \delta J_{i}^{\rm Juno} } \right]^2
  \quad.
  \label{chi_J}
\end{equation}  
$\delta J_{i}^{\rm Juno}$ are the 1-$\sigma$ uncertainties of the measurements. 
Following \citet{MilitzerSaturn2019}, we added the  term $\chi^2_{\rm H-He}$ to generate models with helium rain parameters $P_{\rm rain,1}$ and $P_{\rm rain,2}$ that are compatible with phase diagram of H-He mixtures as derived by \citet{Morales2013}. From the assumed entropy values for the molecular and metallic layers, $S_1$ and $S_2$, one can infer the temperatures $T_1=T(S_1,P_{\rm rain,1})$ and $T_2=T(S_2,P_{\rm rain,2})$ from the isentrope that are given by the equation of state. For the pairs $P_{\rm rain,1}$-$T_1$ and $P_{\rm rain,2}$-$T_2$, we find the closest points on the immiscibility curve by \citet{Morales2013}, $P^*_1$-$T^*_1$ and $P^*_2$-$T^*_2$, and then define
the immiscibility penalty term,
\begin{equation}
  \chi^2_{\rm H-He} = \sum\limits_{i=1}^{2}C_P \left| \frac{P^*_i-P_i}{P_i}  \right| + C_T \left| \frac{T^*_i-T_i}{T_i}  \right| 
  \quad,
  \label{chi_H-He}
\end{equation}  
where $C_P$ and $C_T$ are weights that need to be balanced with those in other $\chi^2$ terms. We set $C_T/C_P=2$. We do not square the individual terms because there is currently no agreement between {\it ab initio} simulation and experimental prediction on the conditions where hydrogen and helium become immiscible. Early path integral simulations showed that the two materials are miscible at 15$\,$000~K \citep{Mi05}. \citet{Vo07} showed with {\em ab initio} simulation that hydrogen and helium are miscible at 8000~K. With more careful {\em ab initio} Gibbs free energy calculations, \citet{Morales2013} predicted hydrogen and helium to phase separate at approximately 6500~K for a pressure of 150~GPa. However, recent shock wave experiments by \citet{Brygoo2021} placed the onset of immiscibility at a much higher temperature of 10$\,$200~K for 150 GPa. Because the deviations of the {\em ab initio} predictions are large and these findings to not yet been reproduced with other laboratory measurements, we will employ the \citet{Morales2013} results when we evaluate the $\chi^2_{\rm H-He}$ term in Eq.~\ref{chi_H-He}.

We also add the wind term,
\begin{equation}
  \chi^2_{\rm wind} = \frac{1}{m} \sum\limits_{i=1}^{m} 
\left\{
\begin{matrix} 
\left[ H(\mu_i)-H_{\rm max}  \right]^2 & {\rm if\,\,} H(\mu_i)>H_{\rm max} \\
0 & \;\;\;\; \;\;\;\; \;\;\;\;\;  {\rm if\,\,} H_{\rm min} \le H(\mu_i) \le H_{\rm max} \\
\left[ H_{\rm min} - H(\mu_i)\right]^2 & {\rm if\,\,} H(\mu_i)<H_{\rm min} \\
\end{matrix}
\right.
\quad,
\label{chi_wind}
\end{equation}  
that keeps the depth of our winds, $H$, within perscribed limits of $H_{\rm min}=1500\,$km and $H_{\rm max}=4500\,$km so that the remain broadly compatible with predictions by \citet{Guillot2018}. We evaluate them at $m=61$ equally spaced $\mu$ points between --1 and +1 with $\mu = \cos(\theta)$ and $\theta$ being the colatitude. We work directly with the observed cloud-level winds from \citet{Tollefson2017} but then make the wind depth to be latitude dependent. One may allow the surface winds to deviate from the observed value and employ the same wind depth for all latitudes. Both types of wind models are compared in \citet{DiluteCore}. 

We solve the thermal wind equation~\citep{Kaspi2016} to derive the density perturbation, $\rho'$,
\begin{equation}
\frac{\partial \rho'}{\partial s} = \frac{2 \omega}{g}\frac{\partial}{\partial z}\left[\rho u\right]
\quad,
\end{equation}
for a rotating, oblate planet~\citep{Cao2017}. Instead of approximating the planet's shape by a sphere, we employ the geometry that is provided by the equi-potential surfaces from the CMS calculation. $z$ is the vertical coordinate that is parallel to the axis of rotation. $s$ is the length of a path that starts form the equatorial plane and follows an equipotential surface that we derived with the CMS method. This method also provides us with the static background density, $\rho$, and the local acceleration, $g$, for a particular model. $u$ is the differential flow velocity with respect to the uniform rotation rate, $\omega$. We represent $u$ as a product of the surface winds, $u_s$, from \citet{Tollefson2017}
and a decay function of $\sin^2(x)$ form \citep{MilitzerSaturn2019}. This function facilitates a rather sharp drop similar to functions employed in \citet{Galanti2021} and \citet{Dietrich2021}. We integrate the density perturbation, $\rho'$, to determine the dynamic contributions to the gravity harmonics before we combine them with the gravity harmonics from the interior, $J_n^{\rm model} = J_n^{\rm interior} + J_n^{\rm wind}$ and enter them into Eq.~\ref{chi_J}. 

Finally we add a number of additional penalty terms,
\begin{equation}
  \chi^2_{\rm guide} = C
\left\{
\begin{matrix} 
\left[ p_{\rm min} - p  \right]^2 & {\rm if\,\,} p < p_{\rm min} \\
0 & {\rm otherwise}\\
\end{matrix}
\right.
\quad,
\label{chi_wind}
\end{equation}  
that help us guide the Monte Carlo ensemble to reach and remain in regions where a certain parameter, $p$, satisfies the condition $p \ge p_{\rm min}$ that we consider physical. Similar terms can assure $p_2 \ge p_1$ for two model parameters. We choose a large value of $C$ like 1000 to assure compliance. We verify that $\chi^2_{\rm guide}=0$ for models that we publish. $Z_1 \ge Z_{\rm protosolar}$ is an obvious condition to satisfy but we also require $Z_2 \ge Z_1$ and $S_2 \ge S_1$. 
\section{Results and Discussion} \label{sec:results}

\subsection{Convergence tests}

\begin{deluxetable*}{crrrrrrrrr}
\tablenum{1}
\tablecaption{Parameters of different interior models in Fig.~\ref{fig:23456}.\label{tab:23456}}
\tablewidth{0pt}
\tablehead{
\colhead{Parameter} & \colhead{2 layer} & \colhead{3 layer} & \colhead{4 layers} &\colhead{4 layers} & \colhead{5 layers} & \colhead{5 layers} & \colhead{5 layers} & \colhead{5 layers} & \colhead{6 layer}
\\
&\colhead{model}&\colhead{model}& \colhead{type A} & \colhead{type B} & \colhead{type A} & \colhead{type B} & \colhead{type C} & \colhead{type D} & \colhead{model} }
\startdata
$Z_1$                               &    0.0206 &    0.0138 &    0.0161 &    0.0166 &    0.0156 &    0.0153 &    0.0153 &    0.0153 &    0.0160  \\
$\alpha$                            &    -      &    -      &    3.00   &    -       &    9.41   &   11.18   &    -      &   10.07   &   13.68	 \\
$P_{\rm   rain,1}$ (GPa)            &  \U{900.00}   &  \U{550.00}   &   93.32   &  \U{442.93}   &   93.08   &   91.64   &  \U{515.00}   &   91.81   &   93.09  	 \\
$P_{\rm   rain,2}$ (GPa)            &  \U{900.00}   &  \U{550.00}   &  760.59   &  \U{442.93}   &  443.17   &  \U{613.90}   &  \U{515.00}   &  636.65   &  609.04 	 \\
$P_{\rm   core,1}$ (GPa)            &  \U{900.00}   &  \U{550.00}   & \U{1189.65}   &  760.90   &  786.17   &  \U{613.90}   &  622.18   &  \U{736.75}   &  731.92	 \\
$P_{\rm   core,2}$ (GPa)            &  \U{900.00}   &  \U{550.00}   & \U{1189.65}   & 2016.15   & 2051.67   &  903.61   &  905.54   &  \U{736.75}   &  916.33 	 \\
$Z_2$                               &    0.1332 &    0.0931 &    0.1639 &    0.1780 &    0.1831 &    0.1095 &    0.1119 &    0.1083 &    0.1178	 \\
$M_{\rm core}^{\rm compact} (M_E)$  &    0.0    &    4.0    &    0.0    &    0.0    &    0.0    &    2.5    &    2.5    &    2.5    &    2.0   
\enddata
\tablecomments{We underlined repeated pressure values because they indicate that a specific layer is not included in a model. Machine readable data files for every model are available online \citet{Zenodo_23456_models}.}
\end{deluxetable*}

\begin{figure}
\gridline{\fig{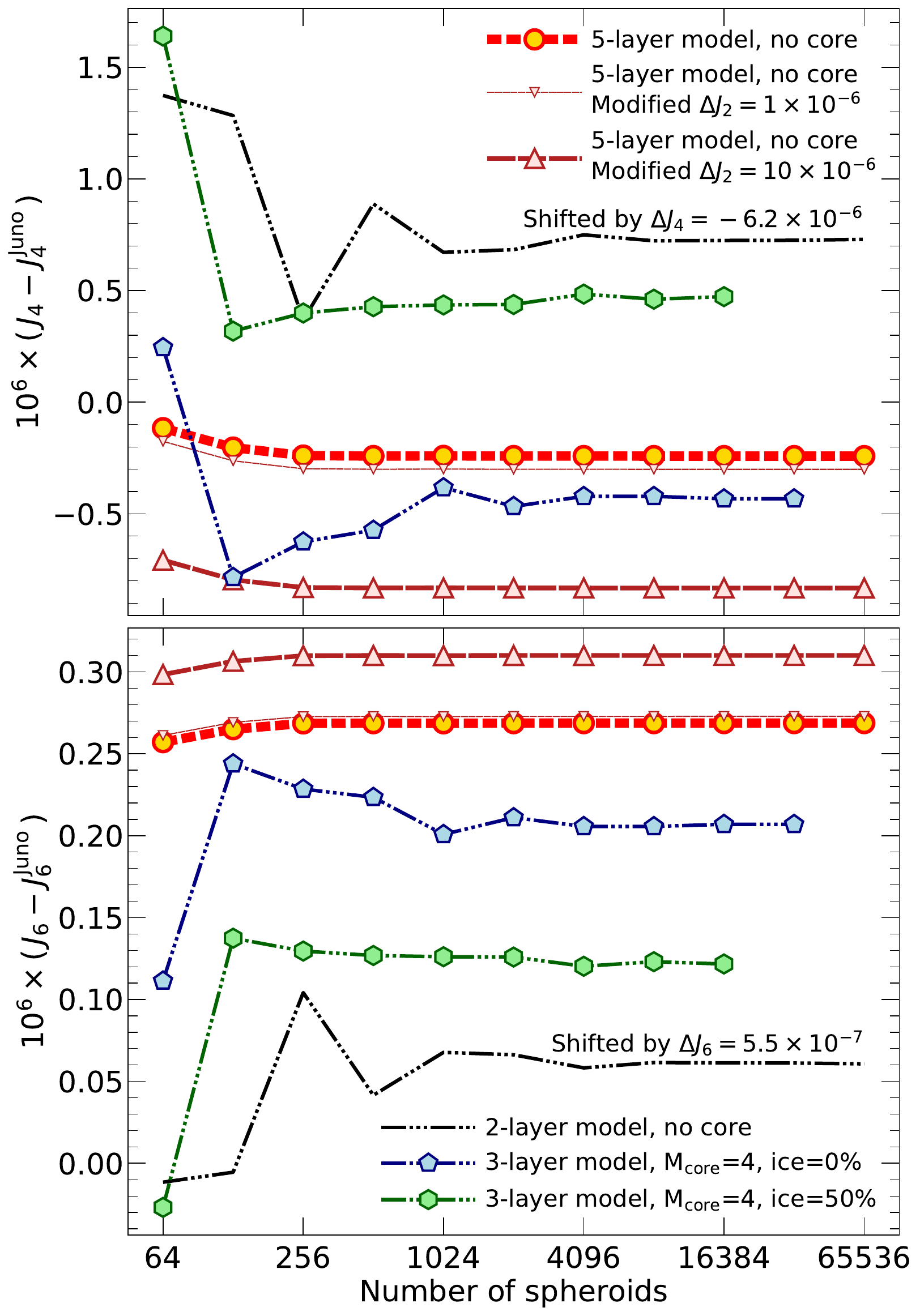}{0.7\textwidth}{}}
\caption{
Convergence of the gravity harmonics $J_4$ and $J_6$ with increasing number of spheroids is demonstrated for selected 2, 3, and 5 layer models that match mass, radius, and $J_2$ as measured by {\em Juno} or a slightly modified value to account for wind effects.  The 3 layer models have compact cores of 4 Earth masses with and without an ice fraction. The 5 layer models do not include any density discontinuities and thus converge much faster than the 2 and 3 layer models. A comparison of the 5 layer models shows that a correction of $J_2$ by 10$^{-6}$ has a small effect on the computed $J_4$ and $J_6$ values. The {\em Juno} measurements have been subtracted from both Y axes. The results of the 2 layer models were shifted to reduce the range of Y axis. All models were constructed without winds.  
\label{fig:Jn}}
\end{figure}

In Fig.~\ref{fig:Jn}, we compare how rapidly selected models of Jupiter's interior converge in $J_4$-$J_6$ space with increasing number of spheroids. All models were constructed to agree with the protosolar helium abundance and to exactly match Jupiter's mass, radius, and, unless noted otherwise, also $J_2$ value as measured by the {\em Juno} spacecraft. The 5 layer models converge the fastest because they do not contain any discontinuities in density. Already with 256 spheroids, the computed $J_6$ value differs only by $2 \times 10^{-10}$ from corresponding result with 65536 spheroids, a difference that is smaller than {\em Juno}'s 1-$\sigma$ error bar of $2 \times 10^{-9}$~\citep{Durante2020}.

Converging $J_4$ is a bit more challenging. With 1024 and 2048 spheroids, $J_4$ differs by $2 \times 10^{-9}$ and $3 \times 10^{-10}$ from the 65536-spheroid result while the {\em Juno} $J_4$ error bar is $8 \times 10^{-10}$. In all following figures, we report results that were obtained with 2048 spheroids that we derived with the accelerated version of CMS method~\citep{MilitzerSaturn2019} using an acceleration factor of $n_{\rm int}=32$. For the models in Fig.~\ref{fig:Jn}, we compared the results derived with $n_{\rm int}=32$ and those of the not-accelerated version ($n_{\rm int}=1$) but the differences were much too small to be discernible in Fig.~\ref{fig:Jn}. This underlines that the convergence of $J_4$ and $J_6$ is foremost controlled by the spheroid number and model type instead of the acceleration factor.

When the coefficients of Jupiter's gravity field are reported~\citep{Durante2020}, the effects of the satellites have been removed from the forces acting on the spacecraft. This includes not only their direct gravitational forces but also the effect of the permanent tides that they introduce in Jupiter via the Love number $k_{20}$. The measured $J_2$ was thus reduced by $6.72 \times 10^{-8}$, a value that was derived with theoretical methods because this effect cannot be measured directly. In Fig.~\ref{fig:Jn}, we include two 5 layer models with altered $J_2$ values in order to determine which alteration magnitude is required to significantly affect the $J_4$ and $J_6$ that we calculated with a specific model. The results in this graph illustrate that $J_2$ modifications of $10^{-6}$ or less are negligible for the purposes of this article, which means the tidal correction of $J_2$ is approximately one order of magnitude too small to matter for our interior models.

In Fig.~\ref{fig:Jn}, we also show 2 and 3 layer models that have an abrupt change in composition in the envelope. Because the density is no longer a continuous function of pressure, many more spheroids are needed to converge $J_4$ and $J_6$ with high precision. With 2048 spheroids, we are able to compute $J_4$ and $J_6$ with a precision of $5 \times 10^{-8}$ and $6 \times 10^{-9}$ or better, respectively. Both values are sufficiently small for the purpose of this study but the convergence could be improved with a more careful treatment of the layer interfaces. 

The magnitudes of $J_4$ and $J_6$ of our 2 layer model in Fig.~\ref{fig:Jn} are smaller than those of our 3 layer models. The 2 layer models better mimic the dilute-core effects in \citet{DiluteCore} than the 3 layer models can because they have 4 Earth masses stored in a compact core, which reduces the dilute-core effect. The magnitudes of $J_4$ and $J_6$ are slightly larger for an ice-free, rocky core composition than for a 50:50 rocky-ice mixture. Ice-free, rocky cores have higher density and the surrounding H-He fluid is thus exposed to a higher pressure, which increases its density and thereby enhances the effect of the core. This is the same trend that we have seen when modeling Saturn's interior \citep{MilitzerSaturn2019}. The masses of rocky cores were found to be slightly smaller than those of rock-ice cores.

\subsection{Two and three layer models}

\begin{figure}
\gridline{\fig{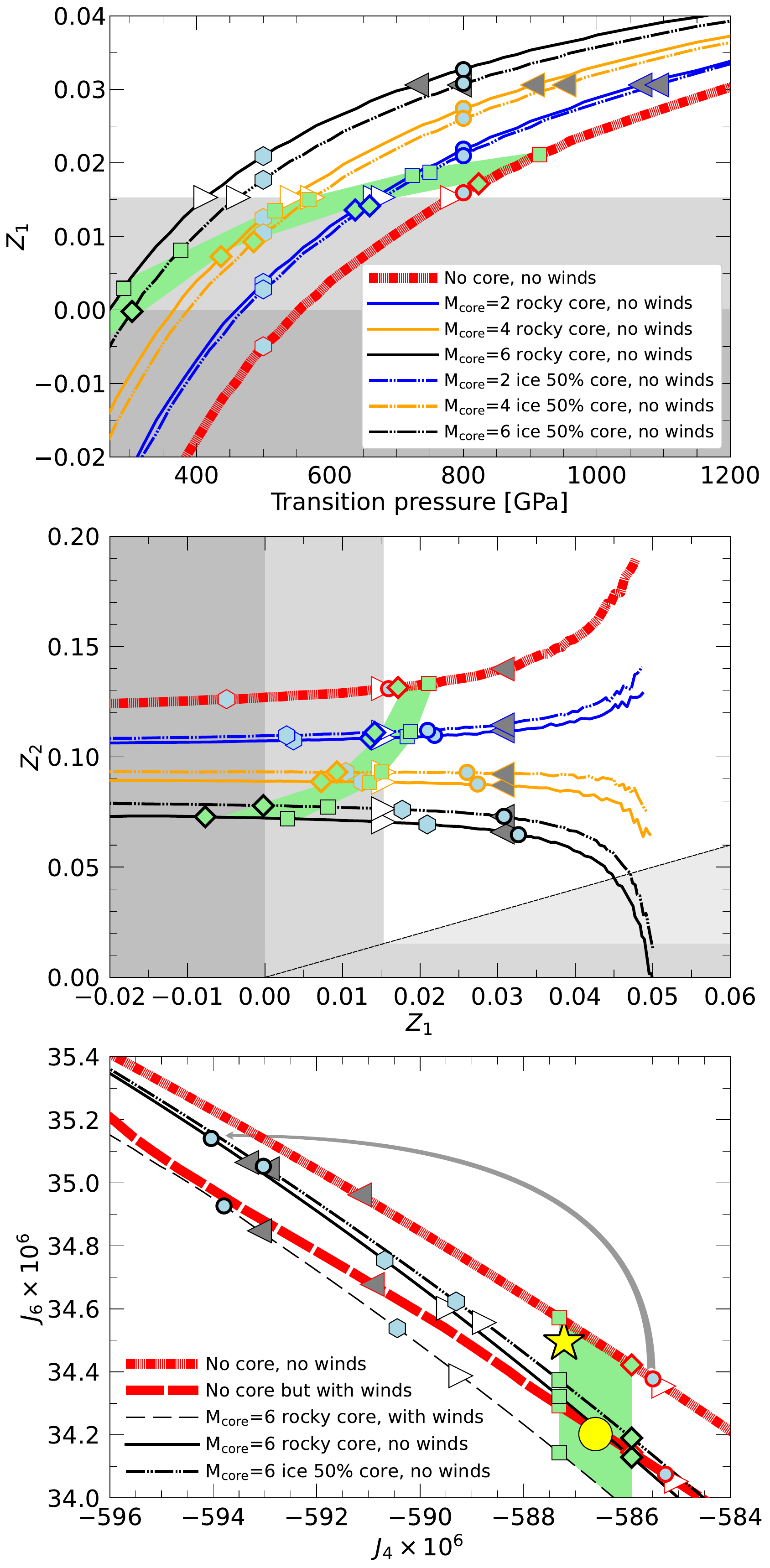}{0.5\textwidth}{}}
\caption{
Predictions from two and three layer models that have a sharp transition in the planet's envelope at a specified transition pressure (x axis of upper diagram). Three layer models have a compact core of 2, 4, or 6 Earth masses (blue, orange, black lines) that is either of rocky (solid lines) and rock-icy (dash-dotted lines) composition. The green symbols and band indicate the range of models with $J_4$ values equal to $J_4^{\rm Juno} \pm 0.7 \times 10^6$ across all panels. The blue filled pentagons and circles indicate models with transition pressures, $P_T$, of 500 and 800 GPa, respectively. The arrow in the lower panel illustrates the shift in $J_4$-$J_6$ space that occurs when a 6 $M_E$ compact core is added a model with $P_T$=800~GPa. The open and filled triangles show models with $Z_1$ values of protosolar or twice protosolar composition. The dark grey region indicates negative and thus unphysical abundances, $Z_1<0$. The medium grey regions show where $Z_1$ or $Z_2$ are less than the protosolar value (0.0153 \citep{Lodders2010}). The shaded triangular region in the middle panel indicates models that are Ledoux unstable ($Z_2<Z_1$) \citep{Ledoux1947}. For reference, the yellow circle in the lower panel shows the {\em Juno} measurements that are exactly reproduced by our preferred five layer models that includes winds. Its prediction from the interior alone are indicated by the yellow star. 
\label{fig:23}}
\end{figure}

The results from our two and three layer models are summarized in Fig.~\ref{fig:23}. All models match mass, radius, $J_2$, and the two conditions for helium abundances $Y_1$ and $Y_2$ so that {\em Galileo} helium measurement is reproduced and the planet overall has a protosolar helium fraction. All models have a sharp transition in the envelope at the prescribed transition pressure, which is their main free parameter. For the three layer models, one can also prescribe the mass of compact core, which can either have a rocky composition with a terrestrial iron fraction of 32.5\% or a rock-ice core with a Callisto-type ice fraction of 50\%. Switching between these two plausible core compositions has only a modest effect on the other inferred properties. The two upper panels of Fig.~\ref{fig:23} show that, for given core mass and transition pressure, ice-free cores increase $Z_1$ and decrease $Z_2$ slightly more than the corresponding rock-ice cores if one compares the predictions with the corresponding core-less two layer models. Also we found that a rock-ice core can have a 13\% larger mass than a ice-free core to yield the same shift in $J_4$-$J_6$ space, which is consistent with predictions for Saturn~\citep{MilitzerSaturn2019} and is driven by the fact that the density contrast between ice and H-He is less than that between rock and H-He. 

The upper two panels of Fig.~\ref{fig:23} shows how $Z_1$ and $Z_2$ respond to variations in the transition pressure. We added symbols and shaded regions to make it easier to follow different trends. The upper panel illustrates that all two and three layer models require a unexpectedly large transition pressure $\gtrsim 400$~GPa for $Z_1$ to be at least of solar abundance. 


The upper panel also shows that, if models are required to match mass, radius, and $J_2$ but not any $J_{n \ge 4}$, an increase in the core mass leads to an increase in $Z_1$. At the same time, the middle panel shows that such an increase of the core mass is accompanied by a reduction $Z_2$ so that the mass in the central region of the planet is approximately preserved. Besides these two main trends, the shape of the curves in the middle panel is controlled by two competing effects. If we increase the transition pressure, the mass of inner layer shrinks, so its heavy element fraction, $Z_2$, may increase, as red curves for core-less models show. Conversely if the inner region already has a sizable compact core ($\gtrsim$ 3 Earth masses), there is less room for heavy elements in the metallic region and $Z_2$ shrinks with increasing $Z_1$ (orange and black curves in the middle panel).

An important constraint on the core mass comes from the {\em Juno} measurements of $J_4$ and $J_6$, which we illustrate in the lower panel of Fig.~\ref{fig:23}. The green symbols and band represent models in the corridor around the {\it Juno's} measurements $J_4$ of size $\pm 0.7 \times 10^6$ that was chosen because it is difficult to construct wind models that yield a larger shift in $J_4$. We replicate the green band in the upper two panels to further constrain the core mass of plausible models. If one combines the $J_4$ interval with the requirement that the outer envelope should have at least a protosolar abundance of heavy elements ($Z_0=0.0153$ \citep{Lodders2010}), models with cores larger than 4 Earth masses can be ruled out. (In  upper two panels, there is no overlap between black curves and the green band in a region that has not been greyed out.) The conditions on $J_4$ (green band) and $Z_1 \ge Z_0$ also require the transition pressure to be 570 GPa or larger, as the upper panel of Fig.~\ref{fig:23} illustrates. Such a high transition pressure is not consistent with the physics of hydrogen-helium mixtures. {\it Ab initio} computer simulations have place the onset of the hydrogen-helium immiscibility region at approximately 100 GPa~\citep{Morales2013}. The discrepancy in pressure makes all three layer models appear less plausible unless one increases Jupiter's interior temperature~\citep{Miguel2022} or changes the EOS to reduce the density of H-He mixtures~\citep{Nettelmann2021}. When \citet{Miguel2022} employed the \citet{MH13} EOS, most models include a helium transition from 300 to 450 GPa (see their Fig.~A1). The trend towards high values is the same that we find here but the pressure values in the work by \citet{Miguel2022} are not as high as ours because they allow the 1~bar temperature to vary and predict values of approximately 183~K. \citet{Miguel2022} also reported a second, smaller group of interior models with a transition pressure of approximately 200~GPa. We do not find models because we do not include large jumps in temperature of up to 2000~K at the transition pressure. 

The bottom panel of Fig.~\ref{fig:23} shows that the inclusion of 6 Earth mass cores leads to large shifts in $J_4$-$J_6$ space away from the {\it Juno} measurements because it diminishes the dilute-core effect that enabled us to match the {\it Juno} measurements in the first place~\citep{DiluteCore}. If one compares rocky-core models without winds and a transition pressure of 800 GPa (blue filled circles), the magnitude of the shift is $\Delta J_4 \approx -8.5 \times 10^{-6}$ and $\Delta J_6 \approx +0.76 \times 10^{-6}$ (see grey arrow). For a lower transition pressure of 500 GPa (blue filled hexagons), the shift is still half as large approximately because the dilute core region contains more mass and the addition of compact core of 6 Earth masses does not have such a large effect. Still it causes the predicted $J_4$ values to fall outside of the green corridor where one could argue that wind models would to able to bridge the remaining difference from the {\em Juno} measurements.



\subsection{Four, five and six layer models}

\begin{figure}

\plotone{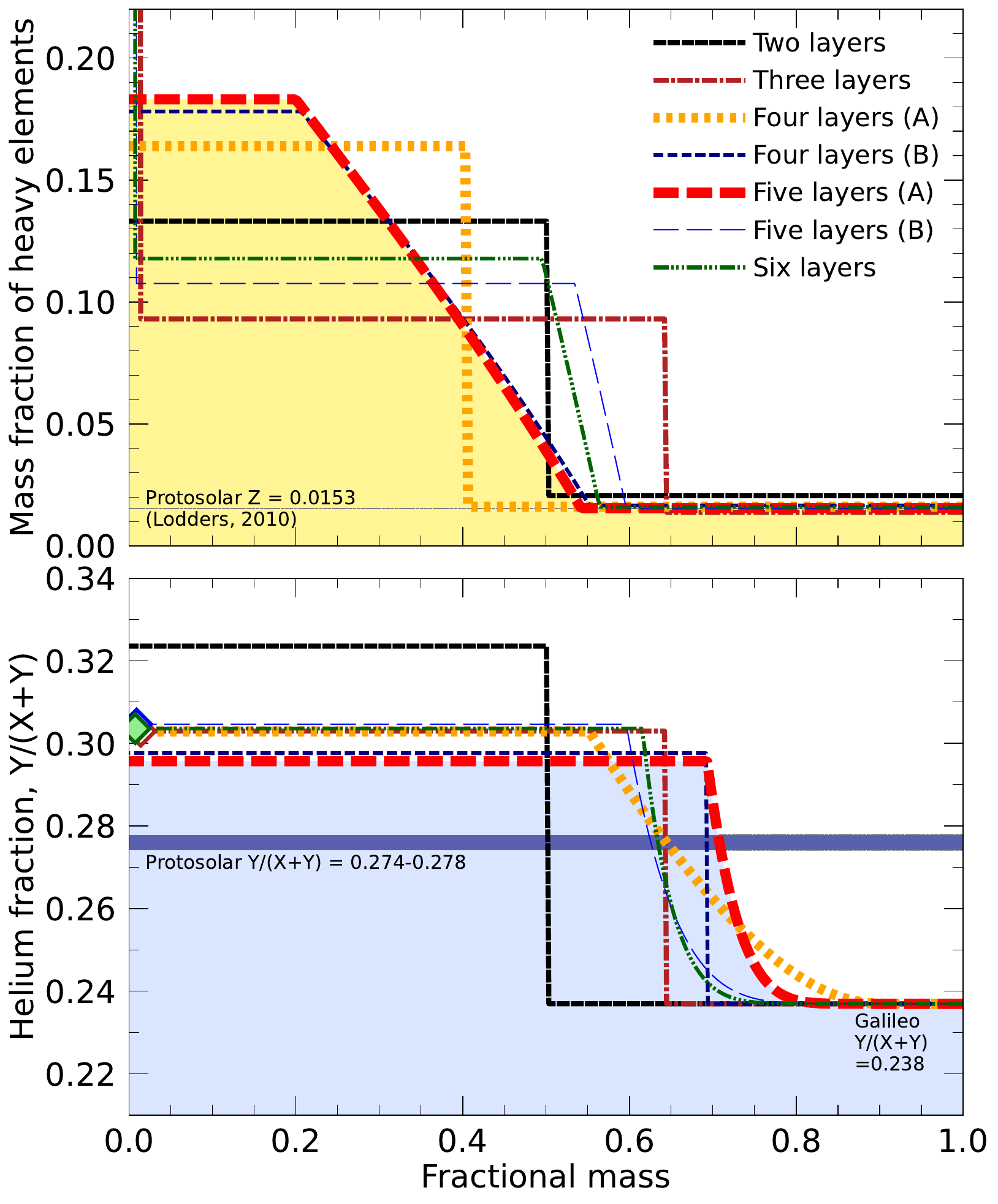}
\caption{
Distributions of heavy elements and helium of models in Fig.~\ref{fig:23456} as function of fractional cumulative mass that has been derived by adding the contribution of different spheroids from the inside out. All models have dilute cores that either end abruptly [models with two, three, and four (A) layers] or gradually [four (B), five (A+B) and six layers]. The three, five (B), and six layer models also have compact cores ($Z=1$) of 4.0, 2.5, and 2.0 Earth masses. In the outer envelope, all models match the helium abundance that the {\em Galileo} entry probe measured. In the interior, they transition to a layer with a higher helium abundance so that the planet has a protosolar hydrogen-helium abundance overall. This transition may either be abrupt [two, three, four (B) layer models] or occur over a extended helium rain region [four (A), five (A+B) and six layer models].
\label{fig:YZ}}
\end{figure}

\begin{figure}
\gridline{\fig{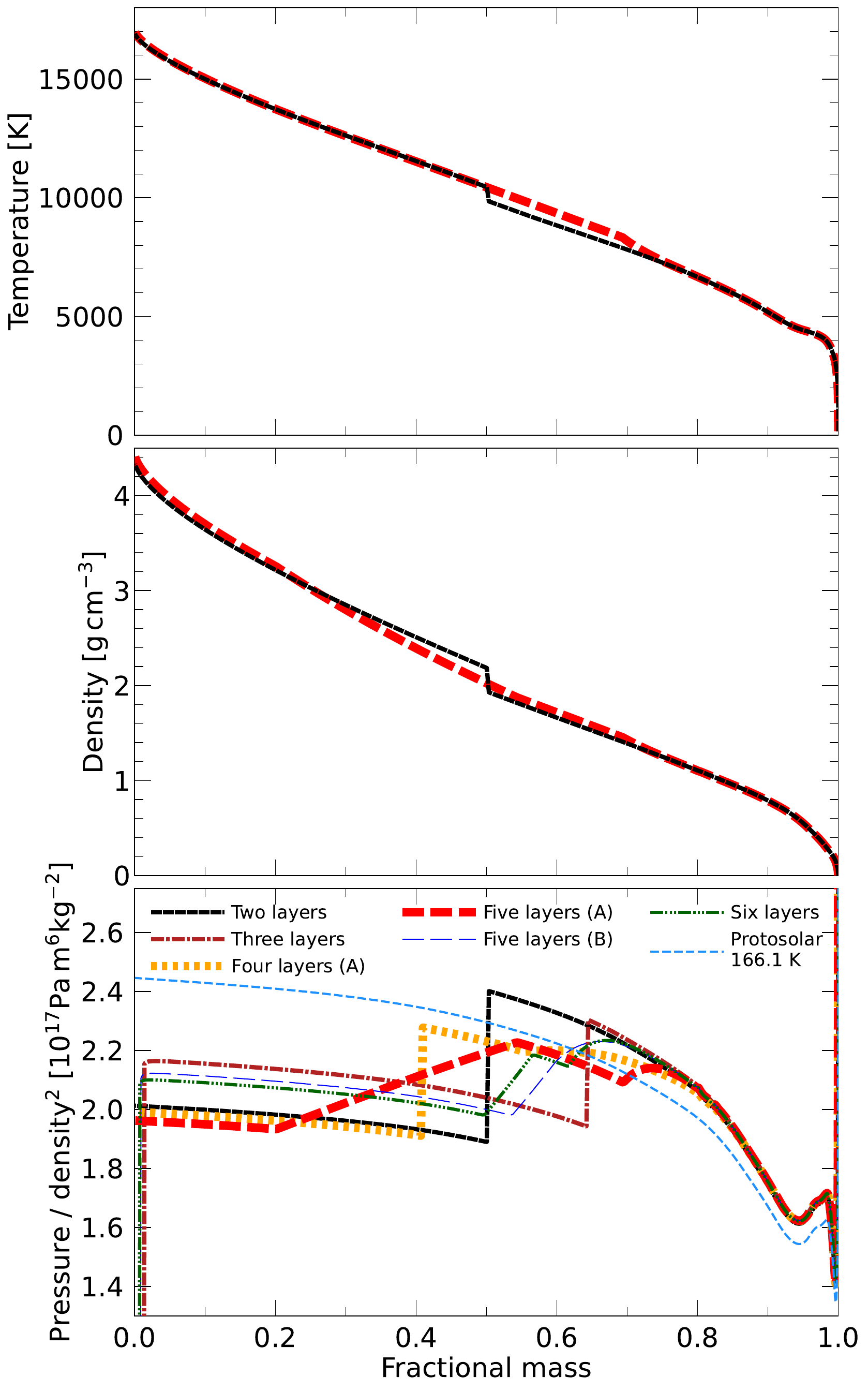}{0.7\textwidth}{}}
\caption{
Temperature, density, and pressure are plotted vs. fractional mass. For clarity only selected models are shown. To better show the differences between models, we scale pressure by density$^{-2}$ because this function is a constant for a polytrope EOS of index 1 \citep{Howard2023}. This scaling is appropriate for the dense part of H-He layers but not for the atmosphere nor inside of a compact core. With increase mass (or radius), the two layer model includes a large decrease in $Z$ at the layer boundary, which decreases its density (middle panel), increases the $P/\rho^2$ term (lower panel) despite an decrease in temperature (upper panel). The $P/\rho^2$ term of our five layer model increases throughout the core transition layer ($M=0.2-0.5$) because $Z$ increases there. The $P/\rho^2$ term is approximately constant inside the dilute core because it is of constant composition. For comparison, we show an isentrope of protosolar composition ($Y_0=0.274$, $Z_0=0.0153$) that starts from a 1~bar temperature of 166.1~K. 
\label{fig:TRhoP}}
\end{figure}

\begin{figure}
\plotone{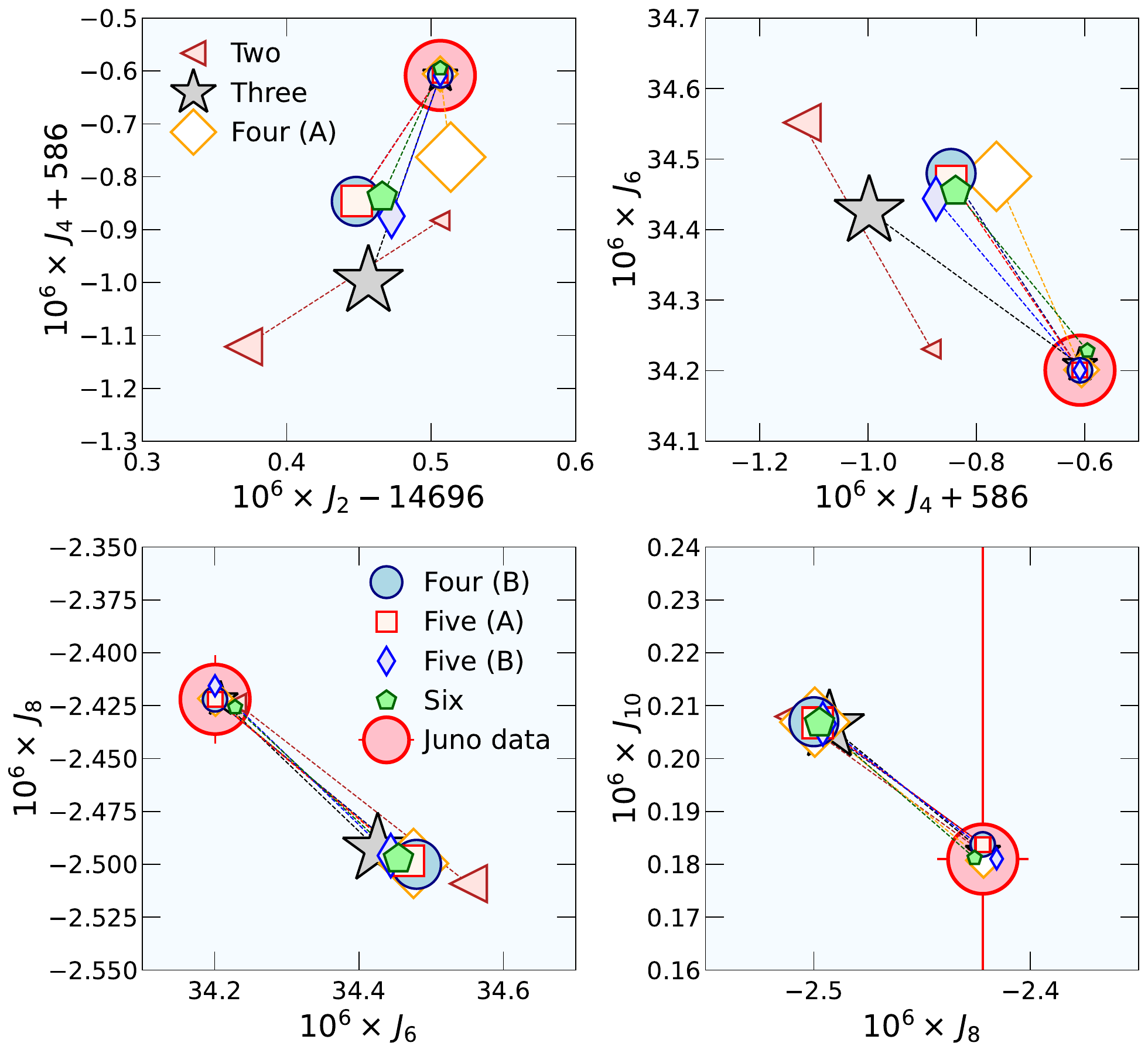}
\caption{
Predictions for the even $J_n$ from different interior models and their
wind contributions are plotted in $J_n$-$J_{n+2}$ spaces.  The {\it Juno} measurements are shown by the large red circles with their 3-$\sigma$ error bars as reported in \citet{Durante2020}. The other symbols represent the seven models from Fig.~\ref{fig:23456}. The large symbols represent the contributions from the interior model only while the small symbols represent interior and wind contributions combined. The combined contributions fit the {\em Juno} measurements very well except for the two layer model that fails to fit $J_4$.
\label{fig:even}}
\end{figure}

In this section, we compare properties of the Jupiter models in Fig.~\ref{fig:23456} that have at least four layers. The corresponding distributions of helium and heavy elements are shown in Fig.~\ref{fig:YZ}. Temperature, density, and pressure are shown in Fig.~\ref{fig:TRhoP}. The temperature profile is determined by an outer isentrope with entropy, $S_1$, where hydrogen occur molecular form, and an inner isentrope with $S_2$, and where hydrogen is metallic. See section \ref{sec:model_assumptions} and \citet{DiluteCore} for details. The bottom panel of Fig.~\ref{fig:TRhoP} includes an isentrope of protosolar composition for comparison. At lower pressures and high fractional mass ($M>0.75$), its density is higher than that of all models because they include a depletion of helium ($Y<Y_0$). At higher pressure and low fraction mass ($M<0.5$), the protosolar density is lower because the models include dilute cores ($Z \gg Z_0$). 

All calculations including wind models were generated under consistent assumptions~\citep{DiluteCore}. The contributions from the interior and winds are illustrated in Fig.~\ref{fig:even}. Except for our two layer models, all other models can fit the {\it Juno} gravity measurements very well. Therefore the following discussion will focus on the question of which models are more plausible given the physics of hydrogen-helium mixtures and, to a lesser degree, on magnetic field information.

With our two layer models, we were not able to match Jupiter's $J_4$ and $J_6$. Even when wind contributions were included, discrepancies of $\Delta J_4 \approx 3 \times 10^{-7}$ and $\Delta J_6 \approx 2 \times 10^{-8}$ remained (see Fig.~\ref{fig:even}). For this reason, we do not discuss the two layer models further. Still, the best models in this class have an abrupt change in composition at $\sim$900~GPa and fractional radius $\sim$0.6.

\begin{figure}
\gridline{\fig{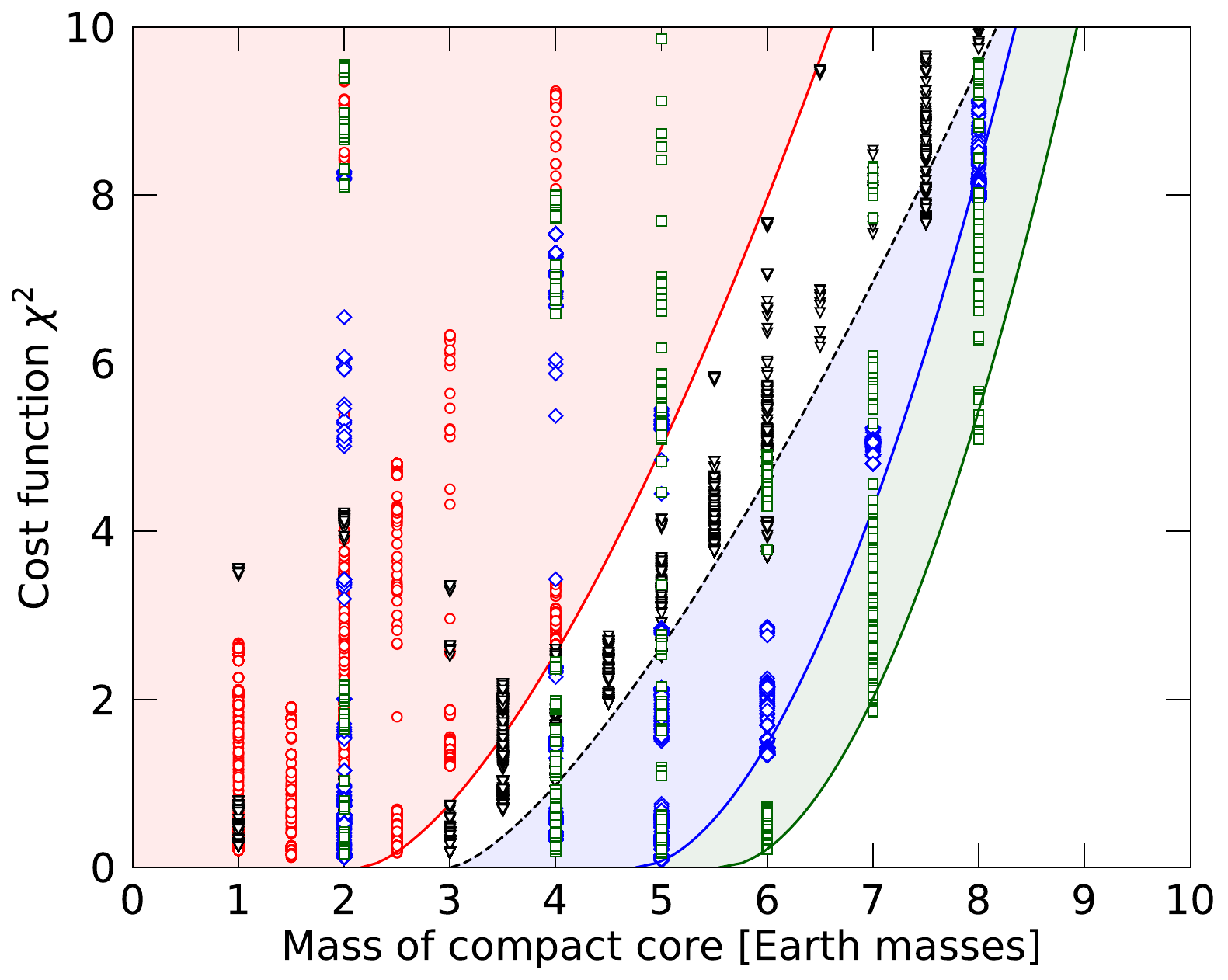}{0.6\textwidth}{}}
\caption{ The symbols show the cost function $\chi^2$ of model ensembles that were constructed with compact cores of a chosen mass under four different assumptions. We added a rocky core to our reference five layer model (the red symbols). The red line and shaded area were added to illustrate that for core masses larger than 2.5 Earth masses, the cost function increases drastically, and models that match the {\it Juno} data well can no longer be constructed. If compact cores of a rock-ice composition are introduced instead, cores of 3.0 Earth masses still yield good models (black dashed lines). If the density is lowered by 2\% over the pressure range from 20 to 100 GPa, compact cores of $\sim$5 Earth masses become possible (blue symbols, line, and shaded area). If the density is lowered by 3\% over the pressure range from 10 to 100 GPa, compact cores of up to $\sim$6 Earth masses still match the {\it Juno} data quite well (green symbols, line, and shaded area).
\label{fig:core_mass}}
\end{figure}

We begin the following discussion with our fiducial five layer model of type A that include a helium-rain layer and a dilute core where the heavy elements only contribute mass fraction of $Z_2\sim0.18$ (see Fig.~\ref{fig:YZ}). Their fraction decreases to $Z_1\approx0.0153$ across the core transition layer where the pressure decreases from 903 to 614~GPa and the fraction cumulative mass increase from 0.20 to 0.54. By inserting a compact core with $Z=1$ into this model, we obtain our six layer models. We are not able to insert a rocky core heavier than 2.5 Earth masses without deviating from the {\it Juno} gravity measurements as we illustrate in Fig.~\ref{fig:core_mass}. If we assume a rock-ice rather than a purely rocky composition for the compact core, the maximum mass increases to 3.0 Earth masses (see Fig.~\ref{fig:core_mass}). Here we assumed the compact core to be homogeneously mixed and to have a Callisto-type composition with 50\%-50\% rock-ice mass fraction. {\it Ab initio} EOS tables were used to derive the density for all materials in the compact core~\citep{WilsonMilitzer2014}. The ice fraction in the compact core reduces its density and therefore decreases the density contrast to the hydrogen-helium mixture, which means the models can accommodate slightly larger compact cores.

It is possible to accommodate larger compact cores but this would require a change in the EOS of hydrogen and helium. If their density were reduced by 2\% below values of the \citet{MH13} EOS in the crucial pressure interval from 20 to 100 GPa, a rocky compact core of $\sim$5 Earth masses could be accommodated. If the density from 10 to 100 GPa is reduced by 3\% compact cores of $\sim$6 Earth masses become plausible as we illustrated in Fig.~\ref{fig:core_mass}. 

The reason for a mass limit on the compact core is that it takes away from the dilute-core effect that enabled us to match $J_4$ and $J_6$ in the first place. Furthermore, Fig.~\ref{fig:23456} and \ref{fig:YZ} show that a compact core of 2.5 Earth masses leads to a reduction of heavy elements in our dilute core from $Z_2\sim$18\%, to $\sim$11\%. More importantly the presence of a compact core increases the dilute core radius to $\sim$0.61 where it decays rather sharply until $R \sim$0.65. It thereby ``pushes'' against the lower boundary of the helium rain layer and thereby shrinks the metallic hydrogen layer. It is consistent with the gravity measurements to eliminate the metallic hydrogen layer entirely, which yields the five layer model of type B in Fig.~\ref{fig:23456}. Alternatively one can keep the metallic hydrogen layer and eliminate the core transition layer to obtain the five layer model of type D. Model types B and D have extended helium rain layers and a compact core, which make them similar to our six layer models. Finally, one  may remove the helium rain layer from our six layer models to obtain our five layer models of type C. To match {\it Juno's} $J_4$ and $J_6$ measurements, the transition to metallic hydrogen must occur at a rather high pressure of $\sim$500 GPa, which makes this model similar to our four layer models of type B. 

Additional information might be harnessed by analyzing Jupiter's magnetic field which was found to be unexpectedly complex~\citep{Moore2018}. Recent magnetohydrodynamic simulations compared the effects of stably stratified (not convective) and fully convective layers in Jupiter's interior and concluded that a stably stratified interior structure would not be compatible with the observed magnetic field~\citep{Moore_2022}. This favors models with thick layers of metallic hydrogen and extended dilute cores because we assume these regions to be homogeneous and convective. 
Still given the magnetic field data, it is not possible to rule out any class of models rigorously. 


There are two obvious choices for generating four layer models from our five layer model of type A. Both are illustrated in Fig.~\ref{fig:23456}. One can either eliminate the core transition layer (model type A) or the helium rain layer (model type B). In A-type models, the dilute core is terminated abruptly ($P_{\rm core,1} = P_{\rm core,2}$) while it decays gradually across the core transition layer of our five layer models. B-type models have a sharp boundary in composition at $P_{\rm rain,1} = P_{\rm rain,2}$ between the layer that is dominated by molecular hydrogen and the layer below that contains mostly metallic hydrogen. 

\begin{figure}
\plotone{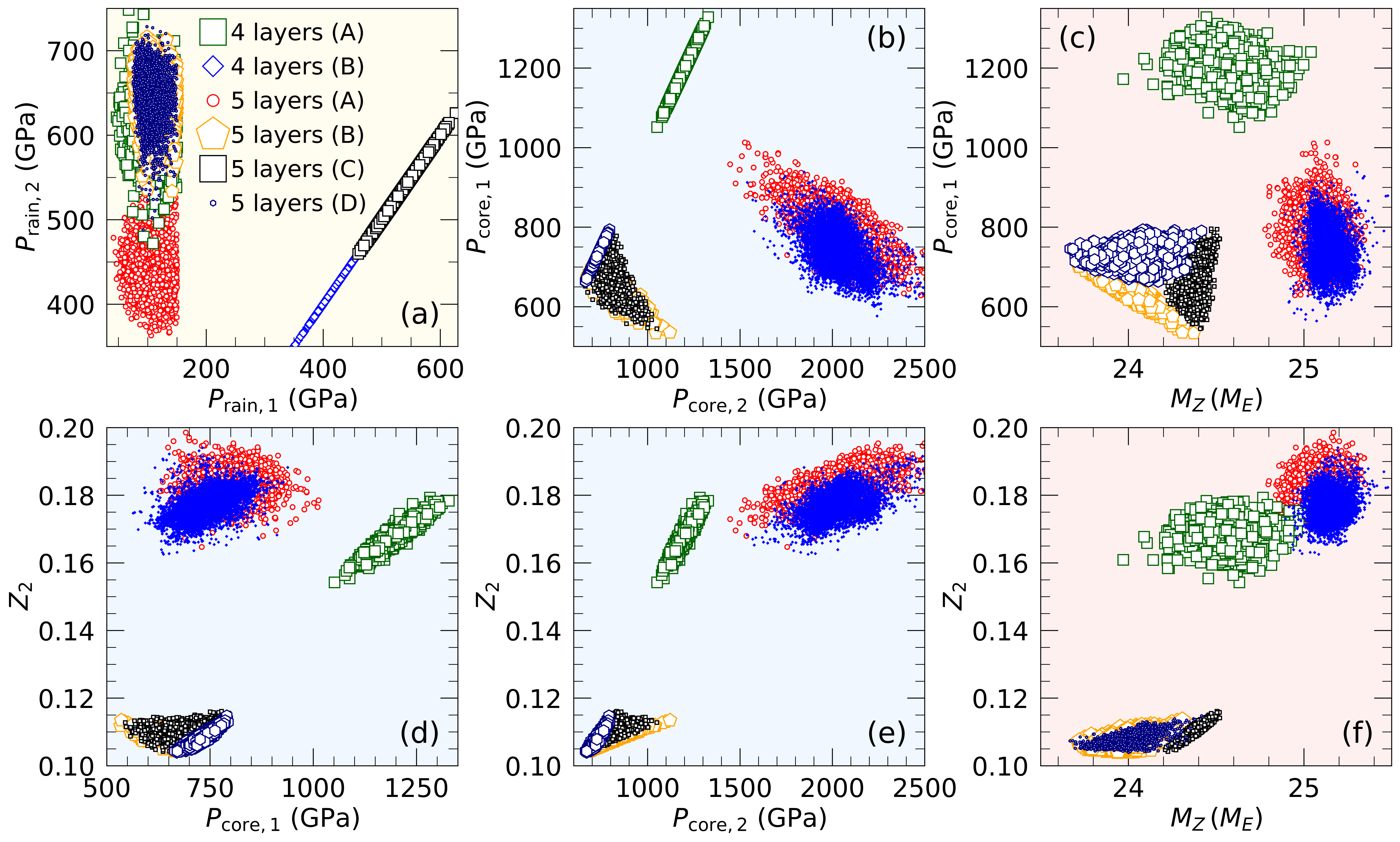}
\caption{ Correlation plot for parameters of our four and five layer models. The four layer models of type B and five layer models of type C have no helium rain layer, which implies $P_{\rm rain,1} = P_{\rm rain,2}$ and leads to the straight diagonal line in panel (a). The other model types plot farther to left in this panel because they have extended helium rain layers with $P_{\rm rain,2} \gg P_{\rm rain,1}$. The four layer models of type A and five layer models of type D have no core transition layer, which implies $P_{\rm core,1} = P_{\rm core,2}$ and means that those models plot on a straight line in panel (b). The other model types have extended core transition layers with $P_{\rm core,2} > P_{\rm core,1}$. Panels (d) and (e) show that the heavy element fraction of the core, $Z_2$, is positively correlated with the outer pressures and in some cases also with the inner pressure of the core transition layer. Panels (c) and (f) show that the total mount of heavy elements varies between 23.5 and 25.5 Earth masses. 
\label{fig:corr}}
\end{figure}

In Fig.~\ref{fig:corr}, we plot correlation among various parameters of our four and five layer models. Our four layer models of type B and five layer models of type C have a sharp instead of gradual change in helium abundance. We find that this change in composition must occur at a rather high pressure from 350 to 550 GPa. This is much higher than a pressure of $\sim$100 GPa for which {\em ab initio} simulations have predicted of onset of the hydrogen-helium immiscibility~\cite{Morales2013}. For that reason, we do not favor Jupiter models that do not include an extended helium rain layer. 

Fig.~\ref{fig:corr} also shows that the helium rain layers of our four layer models of type A and our five layer models of types B and C have more extended helium rain layer than our five layer models of type A (see also Fig.~\ref{fig:23456}). The $P_{\rm core,1}$-$P_{\rm core,2}$ correlation plot shows that our five layer models of type A and four layer models of type B have thick core transition layers that may extend approximately from 650 to 2400 GPa. In our four layer models of type A, the dilute core ends abruptly at $P_{\rm core,1}= P_{\rm core,2}$ from 1000 to 1300~GPa. In our five layer models of type D, this sharp transition occurs at lower pressures from 650 to 800~GPa. Our five layer models of types B and D are not very differ but allow for larger $P_{\rm core,2}$ values of up to 1100~GPa.

Furthermore Fig.~\ref{fig:corr} shows that $Z_2$, the heavy elements abundance of the dilute core, is positively correlated with $P_{\rm core,2}$ and in some cases also with $P_{\rm core,1}$. The reason is that for fixed $Z_2$ value with $Z_2 \gg Z_1$, an increase in $P_{\rm core,2}$ shrinks the dilute core and this leads to a reduction of heavy element abundance of the core region overall (see Fig.~\ref{fig:YZ}). To compensate for such a reduction, $Z_2$ need to rise if $P_{\rm core,2}$ is increased. One also finds that five layer models of types B, C, and D have lower $Z_2$ values because they have larger dilute cores. 

Finally panels (c) and (f) of Fig.~\ref{fig:corr} show that the total amount of heavy Z elements, $M_Z$, in the planet varies between 23.5 and 25.5 Earth masses (7.4--8.0\%). Within the ensemble of a given model type, $M_Z$ is not strongly correlated with $P_{\rm core,1}$ nor with $Z_2$ but it does depend on the model type. Five layer models of types B, C, and D, that include a compact core of 2.5 Earth masses, have a slightly lower total amount of heavy elements of between 23.5 and 24.5 Earth masses. Four layer models of type A, that do not have a core transition layer, have between 24 and 25 Earth masses worth of heavy elements. The highest amounts of between 24.8 and 25.4 Earth mass are predicted for five layer models of type A and for four layer models of type B. Similar results can be expected because both model types differ only by the helium rain layer. It should be noted, however, that one can construct models with substantially higher amounts of heavy Z elements by assuming a superadiabatic temperature exists in the core transition layer \citep{DiluteCore}.

\begin{figure}
\gridline{\fig{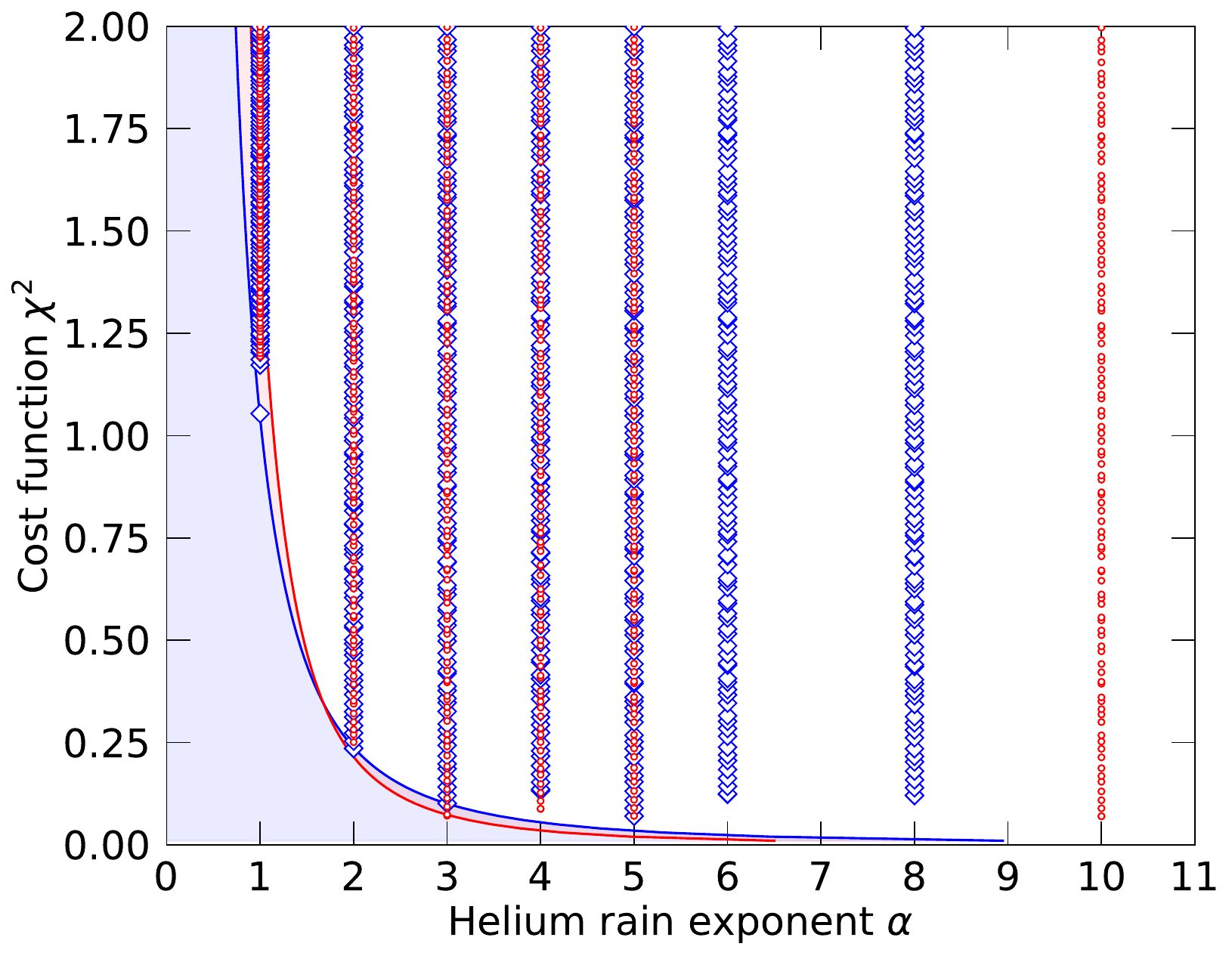}{0.6\textwidth}{}}
\caption{ Cost function $\chi^2$ vs. helium rain parameter, $\alpha$, in Eq.~\ref{exp} is shown for an ensemble of five layer models (blue diamonds) and one of four layer models of type A with a sharp core boundary (red circles). With increasing $\alpha$, more helium has been sequestered to deeper layers, which reduces the density in the upper region of the helium rain layer and thereby makes it easier to construct models that match the {\em Juno} and {\em Galileo} data. For $\alpha$ less than 3, a good match becomes increasingly difficult to obtain, which manifests itself in a sharp rise in the cost function for small $\alpha$ values. This increase is very similar for both type of ensembles as two shaded regions illustrate.     
\label{fig:cRain}}
\end{figure}


Figure~\ref{fig:cRain} was constructed to study whether {\em Juno} gravity measurements allow us to constrain the helium distribution within the helium rain layer. For this purpose, we constructed various MC ensembles for different exponents, $\alpha$ in Eq.~\ref{exp}, for four and five layer models of type A. We find that a linear increase of helium fraction with $\log(P)$ is incompatible with the {\em Juno} measurements and our modeling assumptions. Figure~\ref{fig:cRain} shows that $\alpha$ needs to be three or larger, which means a substantial amount of helium has been sequestered from the upper part of the helium rain layer. 

On the other hand, we were not able to obtain an upper limit of $\alpha$. With increasing $\alpha$, the composition within the helium rain approaches that of the molecular layer above and transition to the metallic layer becomes increasingly sharp. In the limit of large $\alpha$, the five layer models of type A approach the four layer models of type B that require a deep transition ($R\sim0.71$) from 350 to 550 GPa. 

We see the same trend in our three layer models that include a compact core and a sharp transition from molecular to metallic hydrogen. This transition needs to be at $\sim$550 GPa to match the gravity data. This is a pressure value that is too high to be compatible with the physics of hydrogen-helium mixtures. So in conclusion, we do not favor two layer, three layer, B-type four layer models. On the other hand, four layer models of type A, five layer models of types A, B and D but also six layer models with a very modest compact core remain plausible.


\subsection{$J_n$ weight functions}

\begin{figure}
\plotone{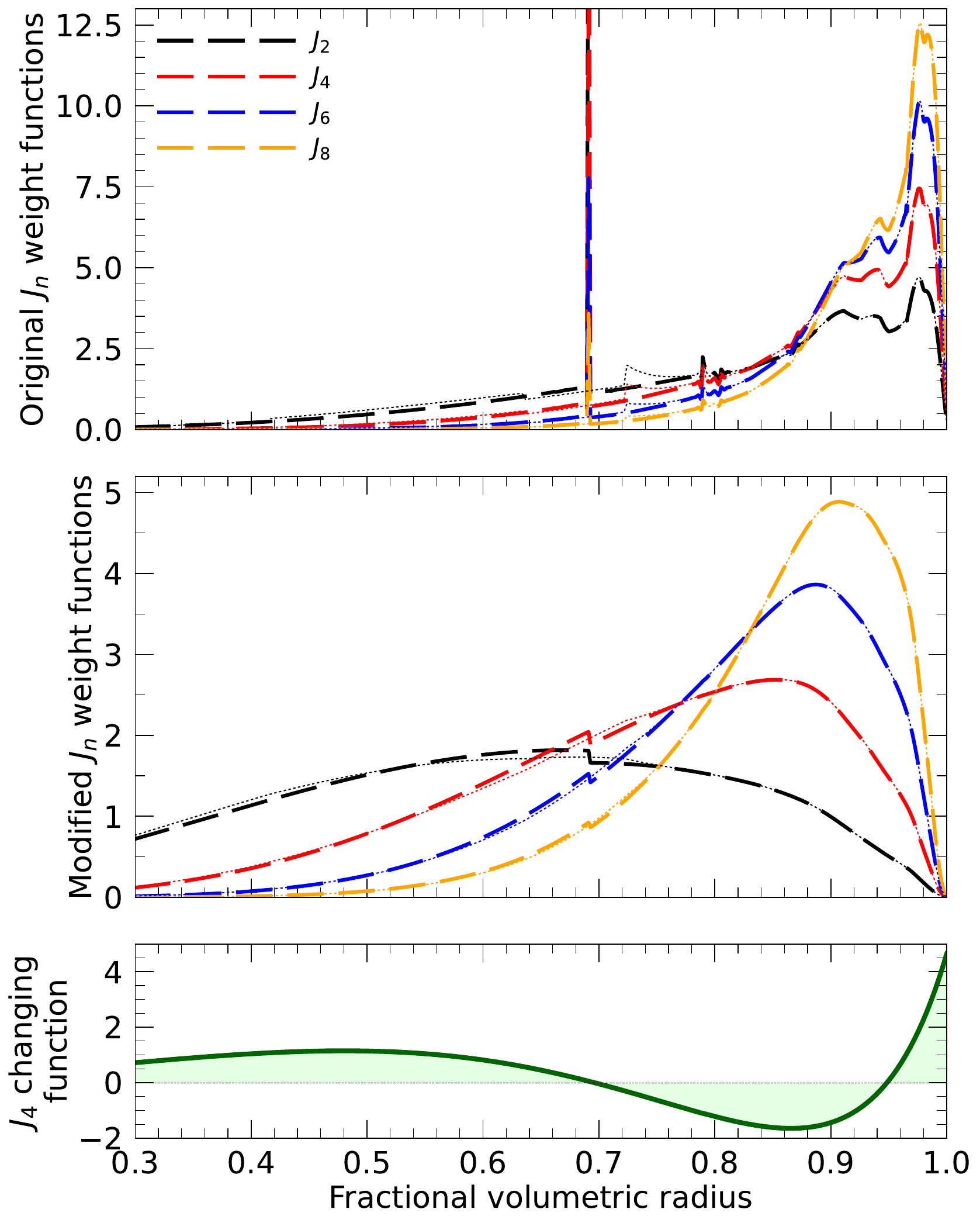}
\caption{
The upper and middle panels show the original and modified weight
functions vs. the volumetric radius. The colors label various even
$J_2$ -- $J_8$. The thick dashed lines show the weight functions for the three layer model
in Figs.~\ref{fig:23456}, \ref{fig:YZ}, and \ref{fig:even} while the
thin dotted lines represent our fiducial five layer model. The lower panel illustrates a density modification in the planet that changes $J_4$ but leaves the planet's mass and $J_2$ unchanged to first order. (All curves remain unchanged regardless whether they were computed with or without acceleration, $n_{\rm int}=32$ and $n_{\rm int}=1$.)
\label{fig:weights}}
\end{figure}

Equation~\ref{weights} specifies which contribution a particular
spheroids makes to the gravity harmonics, $J_n$. Therefore, the
individual terms, $ \lambda_j^n \tilde{J}_{j,n} = \delta_j \hat{J}_{j,n} / M$, have been interpreted as
weights, which have been plotted as function of spheroid radius to
illustrate that the most important contributions to higher order
gravity harmonics come from the outer layers of the
planet~\citep{MilitzerJGR2016}. Here, we propose a slightly different
normalization factor of $1/N_j$ for these functions to remove the
dependence on the choice for the $\lambda$ grid,
\begin{equation}
w^{\rm org}_{j,n} = \frac{ \lambda_j^n \tilde{J}_{j,n} } { N_{j,n} } = \frac{ \delta_j \hat{J}_{j,n} } { M N_{j,n} } \;\;{\rm with}\;\; N_{j,n} = J_n [\lambda_j - \lambda_{j+1}]\;.\label{eq:worg}
\end{equation}
We also divide by $J_n$ to remove their magnitudes from the weight
functions. We show these weight functions in the upper panel of
Fig.~\ref{fig:weights} for our three and five layer models. The
density discontinuity in the three layer model at 550~GPa introduces a
$\delta$ function into the weights at $R \sim 0.7$. From a CMS
perspective, this is very plausible. The discontinuity in density is
represented by one particular spheroid, $j$, whose outer boundary
coincides with the 550~GPa level. This spheroid adds a particularly
large density, $\delta_j$, to the planet, which is equal to the jump
in density at the discontinuity from $\rho_{j-1}$ to
$\rho_j$. Therefore, the weights in Eq.~\ref{eq:worg} resemble in a way
the first derivative of the planet's interior density structure with
respective radius, which explains why the weight functions are not
particularly smooth for $R \gtrsim 0.9$.

When one interprets the weight functions in the upper panel of
Fig.~\ref{fig:weights}, one might be tempted to associate the value of
$w^{\rm org}(\lambda_i)$ with the material that is stored between
$\lambda_j$ and $\lambda_{j+1}$ while in fact it represents an entire spheroid
that has an equatorial radius of $\lambda$ and includes material in
the very center of the planet. However, one can easily derive the
contributions to $J_n$ that arise from material that is stored in
between the spheroids $j$ and $j+1$ and we use this contribution to define a modified weight function,
\begin{equation}
w^{\rm mod}_{j,n} = \frac{ \rho_j }{M N_{j,n}} \left[ \hat{J}_{j,n} -\hat{J}_{j+1,n} \right] \;.
 \label{eq:wmod}
\end{equation}
The original and modified weight functions are normalized in the same way,
\begin{equation}
1 = \sum_j \, [\lambda_j - \lambda_{j+1}] \, w^{\rm org/mod}_{j,n}  \approx \int d\lambda \, w^{\rm org/mod}_n(\lambda)\;. \label{eq:norm}
\end{equation}
We plot these modified weight functions for our three and five layer models in the middle panel of Fig.~\ref{fig:weights}. They still convey the message that the outer parts of the planet contribute most to the higher $J_n$ but now it is easier to see why a dilute core, that extends to $R\sim0.5$, has an impact on $J_4$ and $J_6$. Furthermore the modified weight functions are much smoother than the original ones. A density discontinuity at 550 GPa now only leads to a discontinuity in the weight functions rather than a $\delta$ function. In simple terms one can say that the modified weight functions reflect the contributions from the {\em density} in each layer while the original weight functions represent the contributions from the {\em density increase} in every layer and are therefore closer to core concept of the CMS technique that represents a rotating planet by a sum of nested, constant-density spheroids. The modified weight functions are similar to those that ~\citet{Fortney2016} derived with theory of figures calculations but they are not identical, nor are the weight functions from different theory of figures calculations because assumptions for the interior density structure matter. \citet{N13} predicted the weight functions of $J_2$--$J_8$ for Saturn to respectively have maxima at 0.82, 0.90, 0.92, and 0.94 fractional radii while \citet{Guillot2005} predicted all weight functions for Jupiter to peak within a narrow range of only 0.95--0.97. With CMS calculations for Jupiter, we predict the four weight functions to have maxima at 0.68, 0.83, 0.88, and 0.91 fractional radii respectively.

As illustrated in Eq.~\ref{eq:norm}, one can also view the sum over spheroids as an integral over the equatorial radius $\lambda$. This allows us to rewrite Eq.~\ref{eq:MJ} and ask the question how a small density modification, $\rho'(\lambda)$, changes the computed product $(MJ_n)$,
\begin{equation}
(MJ_n) + (MJ_n)' = \int d\lambda \, \left[\rho(\lambda)+\rho'(\lambda)\right] \tilde{w}_n 
\equiv \left< \rho+\rho',\tilde{w}_n \right> \;\; {\rm with} \;\;
\tilde{w}_n = \frac{\hat{J}_{j,n} - \hat{J}_{j+1,n}} {\lambda_j - \lambda_{j+1}}
\end{equation}
$\left< \ldots \right>$ represents the integral over $\lambda$. The density change can be expressed in terms of $P$ basis functions, $\rho'(\lambda) = \sum_{m=1}^{P} c_m f_m(\lambda)$, which yields,
\begin{equation}
(MJ_n)' = \left< \rho',\tilde{w}_n \right> = \sum_{m=1}^{P} c_m \left< f_m,\tilde{w}_n \right> 
\end{equation}
This linear equation opens up the possibility of adjusting the coefficients, $c_m$, to trigger a change of just one specific $J_n$ while leaving the mass ($M=-J_0$) and other $J_n$ unchanged to first order. In the lower panel of Fig.~\ref{fig:weights}, we give an example for a density modification function that changes $J_4$ but not the mass nor $J_2$. 

This function was derived by setting $P=3$ to consider changes in M, $J_2$, and $J_4$. We opted to introduce $P$ different density modification functions, $\rho_p'(\lambda) = \sum_{m=1}^{P} C_{p,m} f_m(\lambda)$ and to represent their coefficients by the matrix $C_{p,m}$. One can solve for this matrix by inverting the matrix $F_{m,n} \equiv \left< f_m,\tilde{w}_n \right>$ so that every density modification function affects just one of the three variables M, $J_2$, and $J_4$. This requires the matrix product of $\overset\leftrightarrow{C}$ and $\overset\leftrightarrow{F}$ to equal the identity matrix,
\begin{equation}
\overset\leftrightarrow{I} = \overset\leftrightarrow{C} \overset\leftrightarrow{F}
\end{equation}
The function in the lower panel of Fig.~\ref{fig:weights} has positive and negative regions so that mass can be preserved when the density is modified. The function also peaks at large radii where the magnitude of $J_4$ weight function is higher than that of other functions. This is just one specific example. One needs to keep in mind the shape of the resulting density modification functions strongly depends on the choice of basis functions, $f_m$, and on the variable $P$ that specifies how many $J_n$ are considered.

\section{Conclusions}\label{sec:conclusions}

We constructed models for Jupiter's interior with a dilute core that match the recent gravity measurements of the {\em Juno} spacecraft as well as the temperature measurements of the earlier {\em Galileo} entry probe. We employ the nonperturbative CMS method to derive a hydrostatic solution (without winds) for the interior of the rapidly rotating planet and then insert the oblate structure into the thermal wind equation to derive the gravity contributions from the winds. These selfconsistent models allow us matches even and odd harmonics of Jupiter's gravity field under one set of physical assumptions. For example, we use equations of state derived from {\em ab initio} computer simulations to derive the density of given pressure, temperature, and composition. We do not change {\em ab initio} EOS in ad hoc ways, increase the temperature in Jupiter's outer layer nor construct models that have a less than solar abundance of the heavy elements. 

We studied models with different numbers of layers in order to determine which models match the {\em Juno} gravity measurements and identify those that are most compatible with the physics of hydrogen-helium mixture and information on how the a giant planet convects. We found that one cannot match the Jupiter's gravity field with two layer models. A good match is obtained with models that have three layers or more. However, three layer models and four layer models, that have an abrupt transition from molecular to metallic hydrogen, require that this transitions occurs very deep in the planet at $\sim500$ GPa respectively while experiments and {\em ab initio} computer simulations indicate the molecular-to-metallic transition occurs approximately at 100 GPa. For this reason, we find these models less plausible than our four, five and six layer models that include an extended helium rain layer that begins at $\sim$100 GPa. This provides further support for the hypothesis that there are at least two changes in composition in Jupiter's interior: a helium rain layer and a transition to a dilute core.

\citet{Moll2017} studied the time scale for giant planet cores to be eroded convectively under two assumptions. They showed that if there exists only a single sharp boundary in composition between core and the H-He envelope, the core erodes very quickly on a million-year timescale. In this case, the envelope efficiently extracts heat and heavy elements. However, if the core is surrounded by a stably stratified layer, the core will be preserved over billions of years. For this reason we favor our five layer models with a stably stratified core transition layer over our four layer models of type A. They have a sharp core-envelope boundary, which implies the dilute core would be rapidly eroded. 


Finally we derived our six layer models by inserting a compact core into our extended dilute core of our five layer models. If we assume a rocky composition for the compact core, we find it mass cannot be larger than 2.5 Earth masses. If we assume a rock-ice composition this limit increases to 3.0 Earth masses, which is only about 1\% of the planet's mass. Making the compact core more massive than that would require removing so much mass from the dilute core that one can no longer match the gravity harmonics $J_4$ and $J_6$. Based on these arguments, we currently favor our five layer model but a six layer structure with a small compact core of up to 3.0 Earth masses are equally plausible. We expect such models to be revised when the equation of state of hydrogen is measured more precisely, additional {\em ab initio} computer simulations are performed, interior and magnetic field calculations are coupled more tightly or when the {\em June} spacecraft provides us with additional information for the planet's atmosphere or gravitational field.

\appendix

\section{N-Dimensional Root Finder}
\label{RootFinder}

The Newton-Raphson method~\citep{numerical_recipes} is an elegant
procedure to iteratively solve $N$ equations, $\vec{f}(\vec{x})=0$,
with $N$ unknowns that are represented by the vector $\vec{x}$. Starting from
an initial guess, the vector $\vec{x}$ is improved step by step,
\begin{equation} 
\vec{x}^{\rm (new)} =  \vec{x}^{\rm (old)} + \Delta \vec{x} 
\;\;\;\;{\rm with}\;\;\;\; 
\overleftrightarrow{J} \Delta \vec{x} = - \vec{f}\left( \vec{x}^{\rm (old)} \right) \label{Jac}
\;. 
\end{equation} 
This methods requires knowledge of the Jacobian derivative matrix,
$J_{ij}=\frac{\partial f_i}{\partial x_j}$, which is often be
inaccessible in practical applications that involve 
complex functions $\vec{f}$. So here we propose to obtain these
derivatives by linearly fitting the last N+1 points,
$\vec{x}^{(1)}, \ldots, \vec{x}^{(N+1)}$ and the every 
function $f_i(\vec{x})$ separately for $i=1 \ldots N$. To
simplify the following equations, we set $N=2$, but our approach is
general. For every function $i=1$ and 2 separately, we solve this
system of linear equations:
\begin{equation} 
\left[
\begin{array}{cc} x_1^{(1)}-x_1^{(3)}\;\;&\;\; x_2^{(1)}-x_2^{(3)} \\ x_1^{(2)}-x_1^{(3)}\;\;&\;\; x_2^{(2)}-x_2^{(3)} \end{array}
\right]
\left(
\begin{array}{c} \frac{\partial f_i}{\partial x_1} \\ \frac{\partial f_i}{\partial x_2}  \end{array}
\right)
=
\left(
\begin{array}{c} f_i\left(\vec{x}^{(1)}\right) - f_i\left(\vec{x}^{(3)}\right)  \\ f_i\left(\vec{x}^{(2)}\right) - f_i\left(\vec{x}^{(3)}\right)  \end{array}
\right)
\end{equation} 
to obtain approximate values for the Jacobian, $J_{ij}$. Then we perform the
standard Newton-Raphson step in Eq.~\ref{Jac} to obtain the step
$\Delta \vec{x}$ and next point
$\vec{x}^{(4)} = \vec{x}^{(3)} + \Delta \vec{x}$ before dropping the
earliest point, $\vec{x}^{(1)}$. To start, one needs to set a simplex
of $N+1$ points that may be constructed from a single point,
$\vec{x}^{(1)}$, by shifting the $N$ coordinates individually,
$x_j^{(1+i)} = x_j^{(1)} + \delta_{ij} \Delta x_j$ where $\Delta x_j$
represent a chosen step size and $\delta_{ij}$ is the Kronecker delta
symbol. This root-finding algorithm relies in $N+1$ points while the
Newton-Raphson method uses only one. So it is similar to switching
from the 1D Newton method that uses one point $x$ and the derivative
$\frac{df}{dx}$ to the {\it regula falsi}
method~\citep{numerical_recipes} that employs two points but does not
rely on derivatives.

\begin{acknowledgments}
This work was supported by NASA mission {\em Juno} and by the Center for Matter under Extreme Conditions (CMEC) under a grant by the Department of Energy-National Nuclear Security Administration.
\end{acknowledgments}




\end{document}